%% file: shadow.tex
\documentclass[aps,twocolumn,superscriptaddress,prl]{revtex4-2}
\input setup
\input def

\newcommand{\textpo}{PO}
\newcommand{\nPOshadow}{13}
\newcommand{\TPOshadow}{14.8}
\newcommand{\totalRuntime}{5864}

\begin{document}

\title{
Coarse graining the state space of a turbulent flow using
periodic orbits
}
\author{G\"{o}khan Yaln\i z}
\affiliation{IST Austria,
             3400 Klosterneuburg, Austria}
\affiliation{Physics Department,
	Bo\u{g}azi\c{c}i University,
	34342 Istanbul, Turkey}
\author{Bj\"{o}rn Hof}
\affiliation{IST Austria,
             3400 Klosterneuburg, Austria}
\author{Nazmi Burak Budanur}
\affiliation{IST Austria,
	3400 Klosterneuburg, Austria}
\email{burak.budanur@ist.ac.at}

\date{April 29, 2021}

\begin{abstract}
We show that turbulent dynamics that arise in simulations of
the three-dimensional Navier--Stokes equations in a
triply-periodic domain under sinusoidal
forcing can be described as transient visits to the neighborhoods of
unstable time-periodic solutions.
Based on this description, we
reduce the original system with more than $10^5$ degrees of freedom
to a 17-node Markov chain
where each node
corresponds to the neighborhood of a periodic
orbit.
The model accurately reproduces  long-term averages of the
system's observables as weighted sums over the periodic orbits.
\end{abstract}

\maketitle
Producing low-dimensional models of
turbulent flows has been a long-standing scientific challenge with a
wide potential for applications.
Following the discoveries\rf{N90,W98,KawKida01,FE03,WK04,science04} of unstable
time-invariant solutions (equilibria, traveling waves, \ldots)
of three-dimensional ($3D$) fluid flows in pipes and channels,
Gibson \etal{}\rf{GHCW07}
demonstrated the influence of invariant solutions on the dynamics
of plane Couette flow through state space visualizations. Although the
underlying hypothesis that such solutions could eventually be used
for turbulence modeling has been
discussed in subsequent studies\rf{CviGib10,VaKa11,WFSBC15,BDH18},
a clear path towards this goal remained missing.

The studies of invariant solutions of turbulent flows are founded
upon a view of fluid dynamics as a high-dimensional dynamical
system\rf{hopf48}.
In a computational setting, such a dynamical system is constructed
by a spatial discretization that yields a numerical representation
of the fluid's state and a simulator that sets the time-evolution rule.
The simplest invariant solutions of continuous-time
dynamical systems are
equilibria, which in fluid dynamics correspond to velocity fields that
are stationary.
Even though equilibria can influence chaotic flows through
their stable and unstable manifolds\rf{GHCW07,BudHof18},
they by definition lack dynamics and on their own cannot be
used for modeling.
At the focus of the present work are periodic orbits (\textpo{}s),
which form loops in the state space and correspond to velocity fields
that recur exactly after a constant period.

Unstable 
\textpo{}s that are embedded in strange attractors
offer a systematic way of exploring chaos since the
\textpo{}s and
the chaotic trajectories in their vicinity have similar physical
properties\rf{pchaot}.
However, the instability of
\textpo{}s
necessitates special methods for their numerical discovery and
poses a technical challenge especially in high-dimensional
settings such as shear flow turbulence.
Extensive searches for
\textpo{}s in high-dimensional
systems have become
possible after Viswanath's introduction of the
Newton--Krylov--hookstep algorithm\rf{Visw07b}.
Since then, many
\textpo{}s were computed in
plane Couette\rf{CviGib10}
and pipe\rf{WFSBC15} flows where similarities between turbulence and
\textpo{}s were observed.
However, in these studies no attempt was made to
construct a turbulence model based on
\textpo{}s.

In this Letter, we present a quantitatively accurate reduced-order
model of a $3D$ shear flow based on the
numerically computed periodic solutions of the governing equations.
Specifically, we consider $3D$  Kolmogorov
flow\rf{SheWoo1997} under certain symmetry restrictions and
utilize the recently-introduced\rf{YalBud19} state space persistence
analysis for quantifying similarities between turbulence
and
\textpo{}s to show that
the dynamics of this system can be decomposed into
consecutive visits to the neighborhoods of the
\textpo{}s.
Consequently, we propose the neighborhoods
of
\textpo{}s as the bases  of a Markov process that
serves as a coarse-grained model of the turbulent flow.
Upon comparing the long-term observable averages from 
simulations to those obtained from the invariant 
distribution of the
Markov chain, we show
that the 
\textpo{}s give an approximation to the natural
measure\rf{PG97,DasBuch,LaiTel2011} of the system.

$3D$ Kolmogorov flow is described by
the body-forced Navier--Stokes equations
\begin{equation}
    \vfield_t + \vfield \cdot \nabla \vfield
    = - \nabla p + \nu \nabla^2 \vfield + \mathbf{f} \,
    \label{e-Kolmogorov}
\end{equation}
in a rectangular box $[0, L_x] \times [0, L_y] \times [0, L_z]$,
where $\vfield = [u, v, w] (x, y, z)$ and
$p = p(x, y, z)$ are the velocity and pressure fields respectively,
$\nu$ is the kinematic viscosity, 
$\mathbf{f} = \gamma \sin (2\pi y / L_y) \hat{\mathbf{e}}_x$
is the body force with amplitude $\gamma$ 
and 
$\hat{\mathbf{e}}_x$ denotes the
unit vector in the $x$ direction.
$\vfield$ satisfies the incompressibility condition
$\nabla \cdot \vfield = 0$ and
periodic boundary conditions
in all three directions.
The laminar solution of \refeq{e-Kolmogorov} is given by
$u_L = \gamma \nu^{-1} \left(L_y / (2\pi)\right)^2\sin (2 \pi y / L_y)\,, 
v_L = 0\,, w_L = 0$
and it is linearly stable for all $\nu$\rf{vanVeenGoto2016}.
Nevertheless,
turbulence
can be triggered by finite-amplitude
perturbations and is transient at high 
$\nu$\rf{vanVeenGoto2016}. In this sense,
$3D$ Kolmogorov flow admits the
basic phenomenology of the transitional turbulence in
wall-bounded shear flows such as those in pipes and
channels\rf{Manneville2016}.

For numerical integration of \refeq{e-Kolmogorov},
we developed \texttt{dnsbox}\rf{gitdnsbox}, a 
pseudospectral\rf{Orszag1969,Canuto2007} solver
based on the 
\texttt{hit3d} code\rf{Chumakov2007}.
We adapted the Newton--Krylov--hookstep implementation of
\texttt{Openpipeflow}\rf{Willis2017} for finding
\textpo{}s
and utilized
\texttt{scikit-tda}\rf{scikittda2019} for topological data
analysis.
In what follows, we set
$\nu = 0.05$, $\gamma = 1.0$, and
$L_x \times L_y \times L_z = 2 \pi \times 2 \pi \times \pi$. The 
numbers of spatial grid points are
$[N_x, N_y, N_z] = [64, 64, 32]$,
and the second-order predictor-corrector time step is
$\Delta t = 0.0025$.
Fourier-expanded fields are dealiased following the $2/3$
rule and the 
Fourier coefficients show at least four orders
of magnitude drop-off at all times in each direction\rf{Note2}.
The number of nonzero Fourier coefficients after dealiasing
is 110946.
This is an upper bound on the dimension of our system, which, in practice,
is reduced by the divergence-free condition and the imposed symmetries. 
In the supplemental material (SM)\rf{Note2} we provide 
estimates for the effective number of degrees of freedom.

3D Kolmogorov flow is equivariant under
the continuous translations $T_x (\delta x)$ and
$T_z (\delta z)$ in $x$ and $z$ directions by $\delta x$ and
$\delta z$, respectively, and the discrete
symmetries\rf{vanVeenGoto2016}
\begin{eqnarray}
R_{xy} [u, v, w](x, y, z) &=& [-u, -v, w](-x, -y, z) \,, \\
R_y [u, v, w](x, y, z) &=& [u, - v, w](x, -y - L_y /2, z) \,, \\
R_z [u, v, w](x, y, z) &=& [u, v, - w](x, y, - z) \,, \\
S_x [u, v, w](x, y, z) &=& [-u, v, w](- x, y - L_y /2, z)\,.
\end{eqnarray}

We restrict our study to
the flow-invariant subspace of the velocity fields that are
symmetric under $S_x$ and $R_z$, in which
complications due to the continuous
symmetries\rf{BudCvi14} are avoided since only the
translations by
$L_x/2$ and $L_z/2$ in $x$ and $z$ directions 
respectively are allowed. This 
flow-invariant subspace still exhibits transient turbulence with
lifetimes of $O(1000)$, more than $300$ times the
period of our shortest
\textpo{}, i.e.\ the shortest characteristic turnover time.
Since invariance under $S_x$ equates
the action of
$R_{xy}$  and $R_y$,
we can write the symmetry group of the system as
\bea
    G &=& \{\identity, T_{x/ 2}, T_{z / 2}, R_{xy},
    T_{x/2}\, T_{z/2},
    T_{z/2}\, R_{xy}, \continue
    && T_{x/2}\, R_{xy},
    T_{x/2}\, T_{z/2}\, R_{xy}
        \} \, ,
    \label{e-Group}
\eea
where 
$T_{x/2} = T_x (L_x / 2)$,
$T_{z/2} = T_z (L_z / 2)$, and $\identity$ is identity.

The presence of symmetries \refeq{e-Group} implies that each generic
state of the system has $7$ symmetry copies.
Since our analyses require parsing large data sets, it is
crucial to eliminate redundancies in the data.
With this in mind, we construct a
symmetry-reduced representation of
our system 
via a state space coordinate transformation.
Let $\tilde{\svec}$ be a state vector holding the real and imaginary
parts of coefficients in the Fourier expansion of $\vfield$.
Noting that each element of \refeq{e-Group} is its own inverse,
we decompose $\tilde{\svec}$ into symmetric and antisymmetric components
under the action of $\sigma \in G$
as
$\tilde{\svec}^{\pm}_{\sigma}
=\frac{1}{\sqrt{2}} (\identity \pm \sigma) \tilde{\svec}$.
By construction under the action of $\sigma$,
the elements of $\tilde{\svec}^{+}_{\sigma}$ are
invariant
and those of $\tilde{\svec}^{-}_{\sigma}$ change
signs.
Let $(\rho_1, \rho_2, \rho_3, \rho_4, \ldots)$ be the elements
of $\tilde{\svec}^{-}_{\sigma}$, we write the invariants of
$\sigma$ as
\beq
\left\{\frac{\rho_1^2 - \rho_2^2}{\sqrt{\rho_1^2 + \rho_2^2}},
\frac{\rho_1 \rho_2}{\sqrt{\rho_1^2 + \rho_2^2}},
\frac{\rho_2 \rho_3}{\sqrt{\rho_2^2 + \rho_3^2}},
\frac{\rho_3 \rho_4}{\sqrt{\rho_3^2 + \rho_4^2}},
\ldots
\right\}\,. \label{e-Invariants}
\eeq
These invariants, without the denominators, were written for
the Kuramoto--Sivashinsky system in \refref{BudCvi15}.
Here, we introduce the denominators
to prevent the transformation from producing numbers that are
too large or small.
One can confirm by inspection that
the elements of \refeq{e-Invariants} are invariant when all
$\rho_i$ change their signs but not when any other
subset of $\rho_i$ does.
Thus, replacing the elements of $\tilde{\svec}^{-}_{\sigma}$ with
\refeq{e-Invariants} gives us coordinates that are invariant under
$\sigma$.
We begin this procedure with the reduction of $T_x (L_x/2)$,
and repeat for $T_z (L_z / 2)$ and $R_{xy}$ 
to obtain
the 8-to-1 transformation to
the symmetry-reduced coordinates $\svec$.

At the first stage of our study, similar to
\refrefs{CviGib10,WFSBC15},
we generated turbulent data sets 
from random initial conditions with a total run time of \totalRuntime{} 
and initiated Newton--Krylov--hookstep searches for
\textpo{}s
from near-recurrences
of the turbulent flow as measured by
$R(\zeit, \zeit') = \| \vfield (\zeit + \zeit') - \vfield(\zeit)\| / \| \vfield(\zeit) \|$,
where $\| \vfield \|^2 = \frac{1}{L_x L_y L_z} \int \vfield \cdot \vfield\ d\mathbf{x}$.
With the choices of recurrence threshold $R_{th} = 0.3$ 
for triggering 
\textpo{} searches
and recurrence time $\zeit' \in [0, 20]$,
this process
resulted in $18$ distinct
\textpo{}s with relative errors
$\| \vfield_p(t+T_p) - \vfield_p(t) \| / \| \vfield_p(t) \|$ less than
$10^{-9}$.
We found two of these 
\textpo{}s to have very similar 
physical properties and thus discarded one of them to retain 
17.
This omission had no significant 
effect on our results\rf{Note2}.
Hereafter, we refer to these orbits as $\po_i$ with
indices $i = 1, 2, \ldots, 17$ ordered in
increasing periods, where the shortest period $T_1 = 2.8$ and the
longest one $T_{17} = 17.3$\rf{Note2}.

\begin{figure}[h]
\begin{overpic}[height=0.31\linewidth]{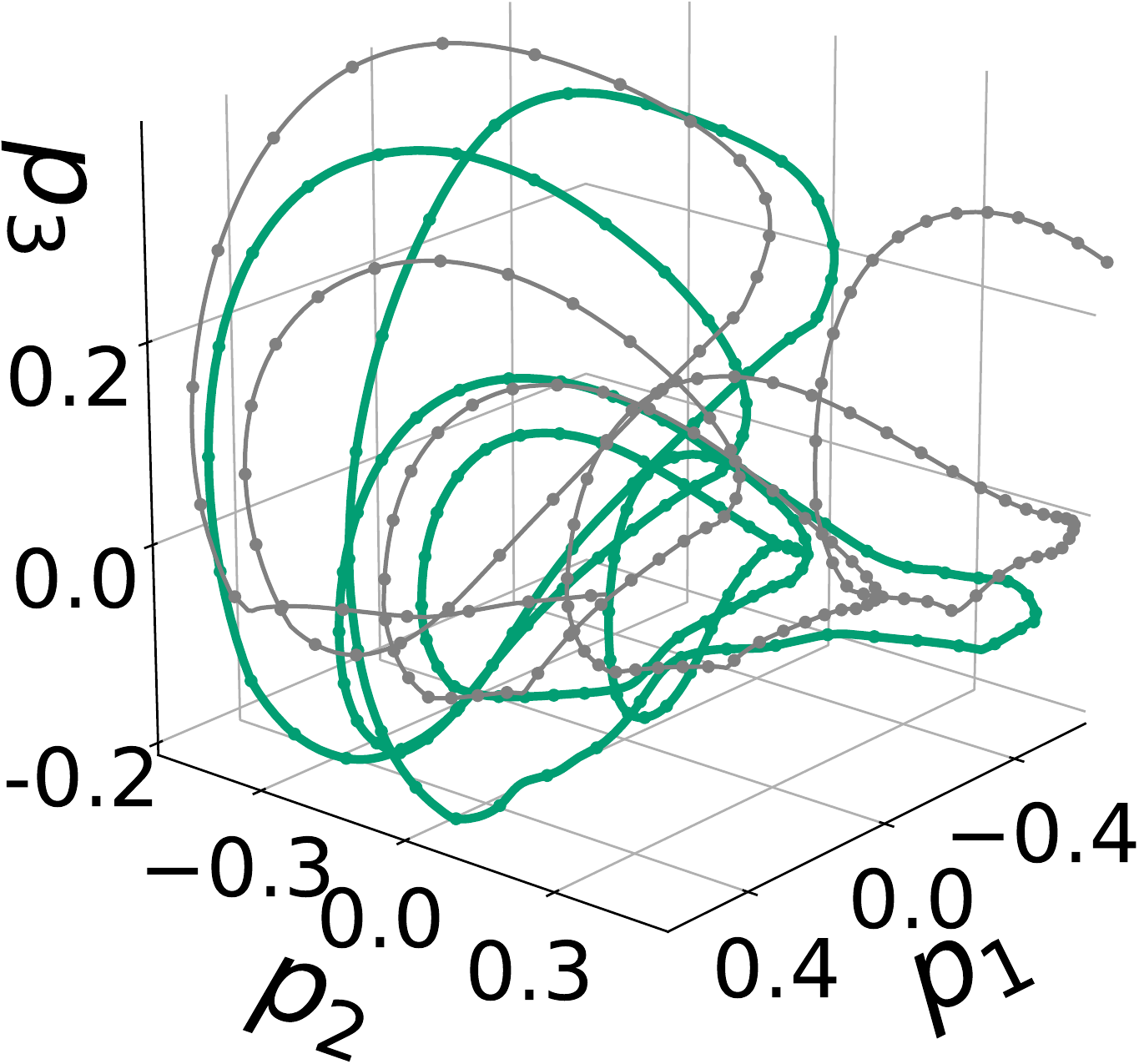}
    \put (0,0) {(a)}
\end{overpic}
\begin{overpic}[height=0.31\linewidth]{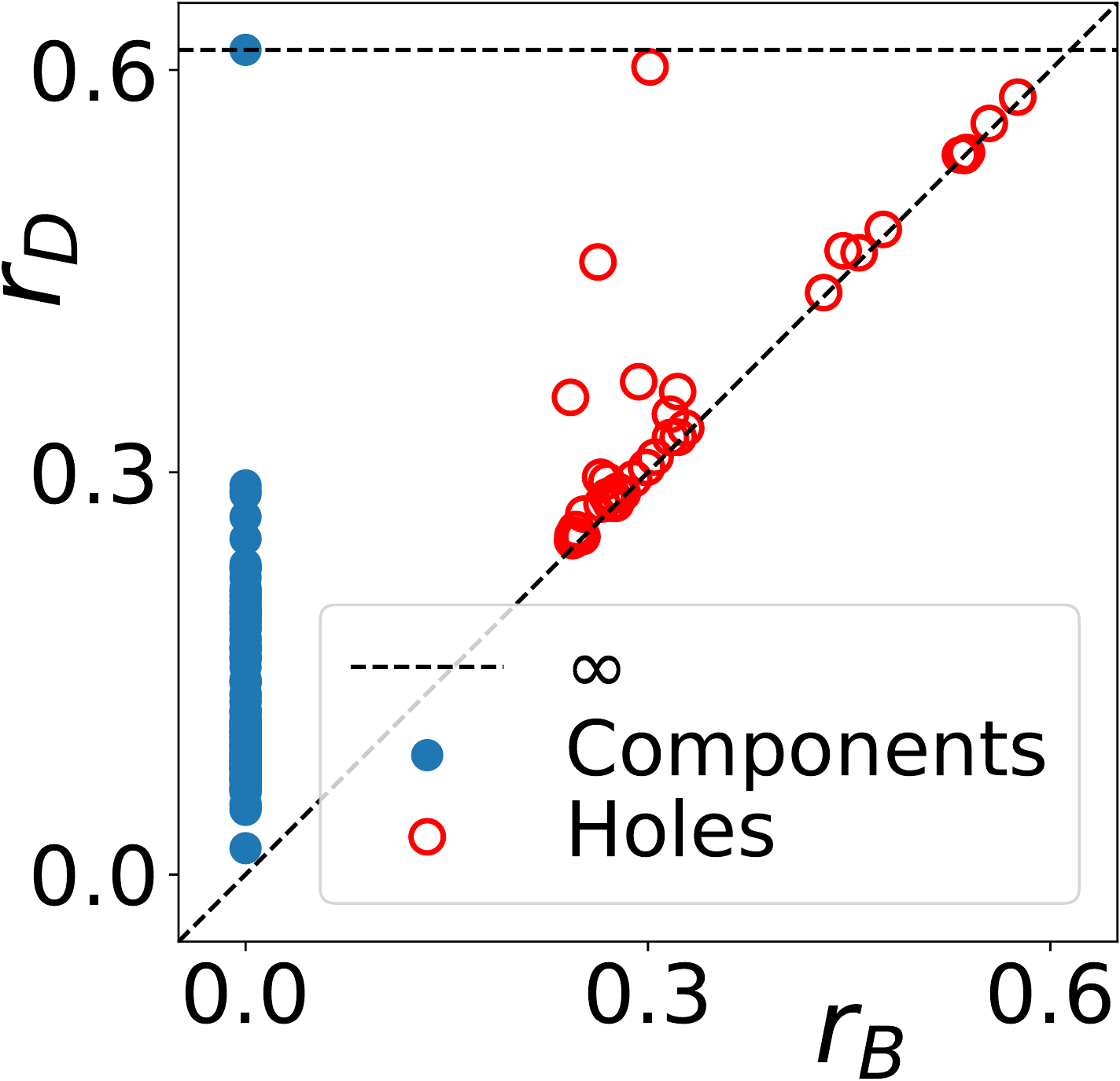}
    \put (0,0) {(b)}
\end{overpic}
\begin{overpic}[height=0.31\linewidth]{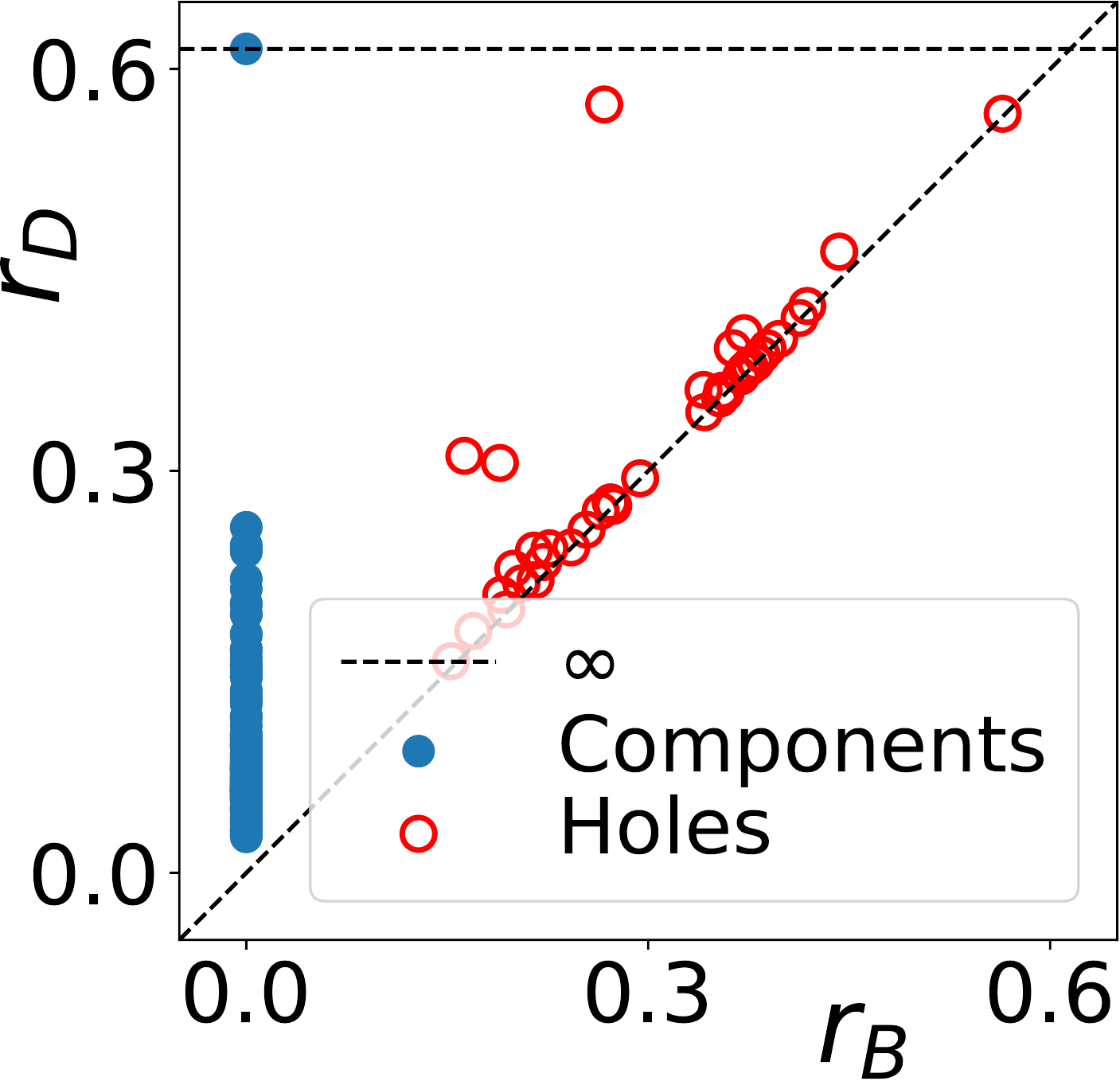}
    \put (0,0) {(c)}
\end{overpic}
    \caption{(Color online)
        (a)
        $\po_{\nPOshadow}$ (green/thick)
        and a shadowing trajectory (gray/thin) visualized as
        projections onto the leading three principal components
        of $\po_{\nPOshadow}$.
        (b,c) The persistence diagrams associated with
        $\po_{\nPOshadow}$ (b) and shadowing trajectory segment (c)
        shown in (a).
        The data points used for generating the
        persistence diagrams (b,c) are marked with dots along
        the projection curves in (a).
        \label{f-ShadowingPD}}
\end{figure}

The first question that we address is how frequently
individual 
\textpo{}s are visited, \ie\ \emph{shadowed}, by
the turbulent flow.
Our analysis begins with producing projection bases 
for individual
\textpo{}s. To this end,
we take snapshots along one period of each orbit with the sampling
time $t_s = 0.1$ and generate the associated principal components\rf{Jolliffe2002}
in the symmetry-reduced state space using the $L_2$ inner product 
$\inprod{\svec^{(k)}}{\svec^{(l)}} = \sum_i \svec^{(k)}_i \svec^{(l)}_i$.
Next, we simulate turbulent flow and project it onto each
of these bases centered at the empirical mean of the respective 
\textpo{}.
As an illustration, \reffig{f-ShadowingPD}(a) shows
$\po_{\nPOshadow}$ along with a shadowing turbulent trajectory spanning a time
window equal to the period $T_{\nPOshadow} = \TPOshadow$ of $\po_{\nPOshadow}$
as projections
onto the leading three principal components of $\po_{\nPOshadow}$.
The main idea of state space persistence analysis\rf{YalBud19} is quantifying
the shape similarity of projections of the
\textpo{}s and those
of turbulent trajectories such as the ones shown in
\reffig{f-ShadowingPD}(a).
For this purpose, we utilize persistent homology, which we briefly describe next
and refer to
\refrefs{EH2008,ECE2011,Otter2017} for in-depth introductions.

Persistent homology is a computational topology
method for extracting shape information from a data set by generating its
representations at different resolutions and tracking the topological changes
in the process.
In our applications, the data sets of interest are the state space projections
such as those visualized
in \reffig{f-ShadowingPD}(a) and the final products 
of the persistence computation are the \emph{persistence diagrams},
examples of which are shown in  \reffig{f-ShadowingPD}(b,c).
Each marker in a persistence diagram corresponds to the
\emph{birth} and \emph{death} of a topological feature represented
by the pair $(\resol_B, \resol_D)$ of birth and death resolutions.
For every data set $\DataSet$, persistent homology gives us two
diagrams $\PerD_0$ and $\PerD_1$ corresponding to the
components and holes, respectively~\footnote{
In general, higher-dimensional voids are included in persistence
diagrams, however, in our applications we consider
one-dimensional holes only.}.
What is gained in this process is  a way of quantifying the shape
similarity since one can define a metric in the space of
persistence diagrams.
Assuming each diagram also contains the trivial elements at
the diagonal $r_B = r_D$ with infinite multiplicity,
we can define
the \emph{bottleneck distance} between
$\PerD^{(k)}$ and $\PerD^{(l)}$ as
\begin{equation}
    W_{\infty} (\PerD^{(k)}, \PerD^{(l)}) =
    \inf_{\phi}
    \sup_{\mu \in \PerD^{(k)}}
    ||\mu -  \phi(\mu)||_{\infty} \, ,
    \label{e-bottleneck}
\end{equation}
where $\phi: \PerD^{(k)} \rightarrow \PerD^{(l)}$ is a bijection
from $\PerD^{(k)}$ to $\PerD^{(l)}$.
The bottleneck distance \refeq{e-bottleneck} can be interpreted as
the largest (measured in the $L_\infty$ norm)
of the shortest one-to-one pairings of the elements of
$\PerD^{(k)}$ and $\PerD^{(l)}$.
An important property of persistent homology that motivates
our application is \emph{stability}\rf{CEH2007}:
Small perturbations to the underlying data result in small
variations, measured by the bottleneck distance \refeq{e-bottleneck}, of
the associated persistence diagrams.

\begin{figure}
\begin{overpic}[width=0.99\linewidth]{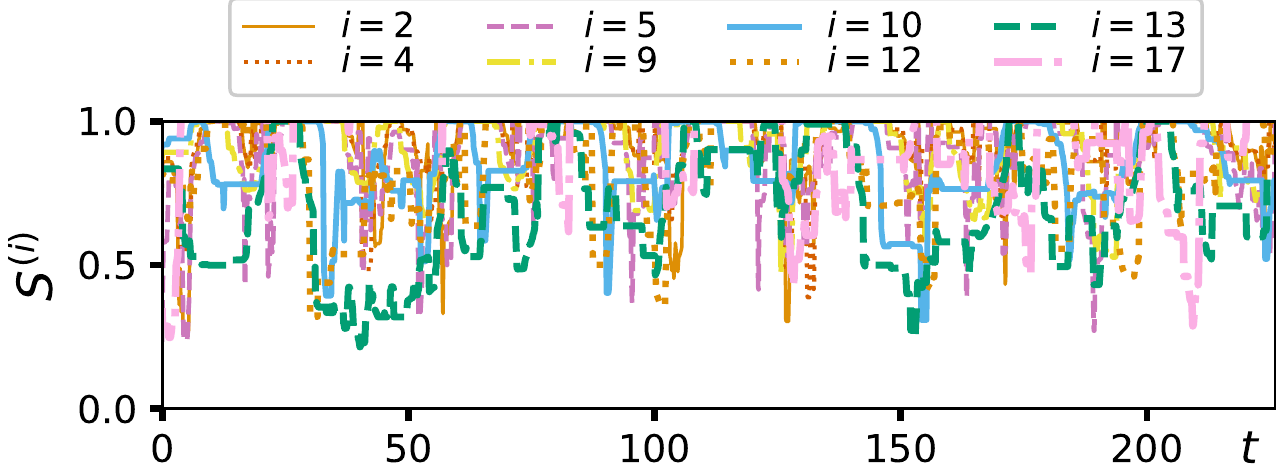} 
    \put (0,-2.5) {(a)}
\end{overpic}\\
\vspace{0.3cm}
\begin{overpic}[width=0.99\linewidth]{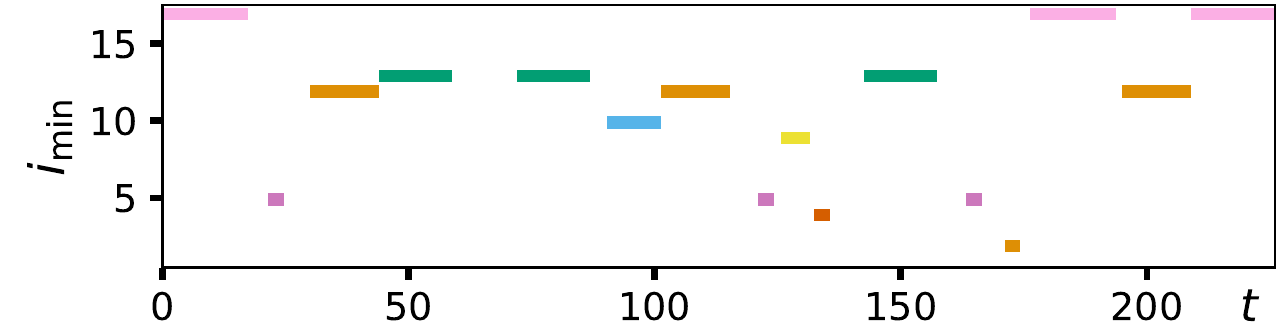} 
    \put (0,0) {(b)}
\end{overpic}\\
\vspace{0.1cm}
\begin{overpic}[width=0.99\linewidth]{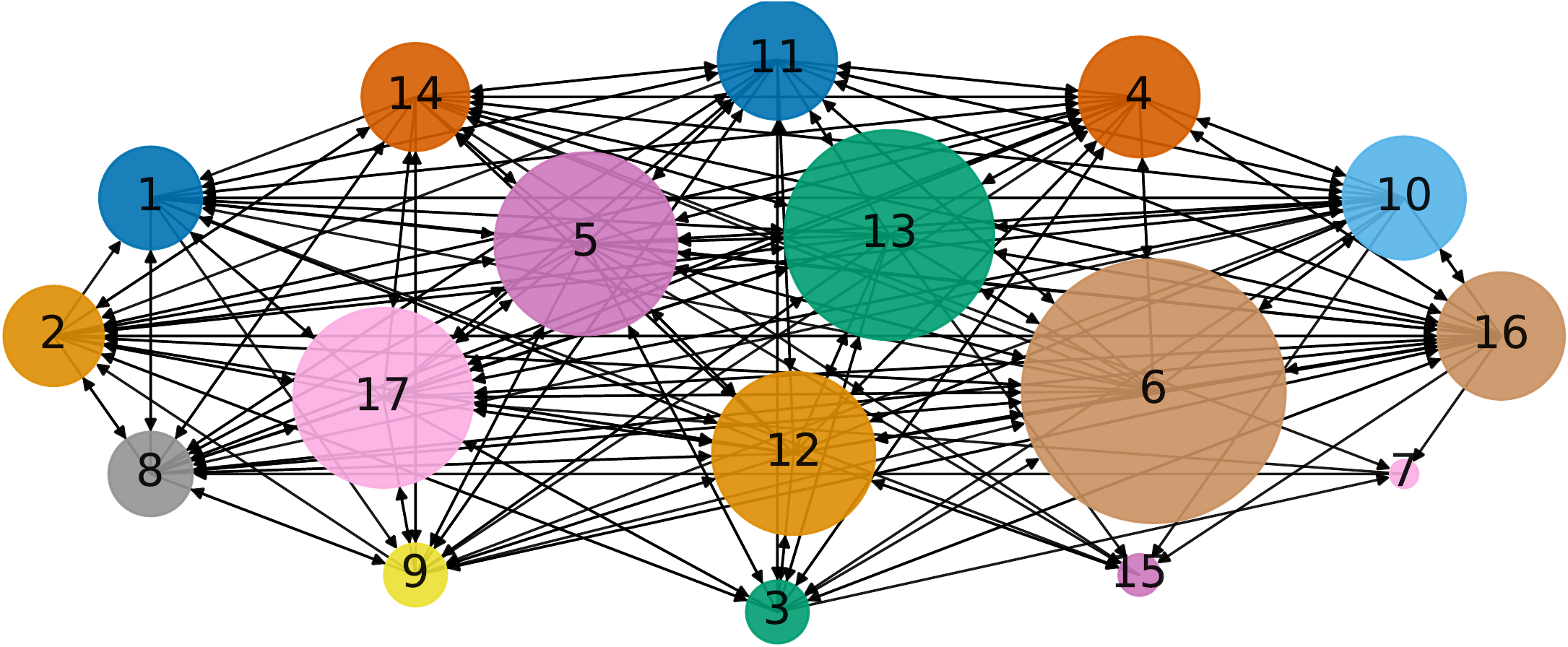} 
    \put (-0.01,0) {(c)}
\end{overpic}
    \caption{(Color online)
        (a)
        Shadowing distance of a turbulent trajectory from
        $8$
        \textpo{}s.
        (b)
        Shadowing decomposition of the turbulent
        trajectory.
        (c)
        State transition graph where the nodes correspond to			
        \textpo{}s and
        arrows indicate the
        possible transitions between them.
        The self-loops are omitted for clarity and
        the node sizes are proportional to the probability
        of observing the respective
        \textpo{} as inferred
        from the invariant distribution of the corresponding
        Markov process.
        \label{f-ShadowDecomp}
    }
\end{figure}

We are now in position to define the \emph{shadowing distance}.
Let
$\DataSet^{\po_i} = \{\svecproj^{\po_i} (0),
                        \svecproj^{\po_i} (t_s),
                        \ldots,
                        \svecproj^{\po_i} ((N_i - 1)t_s) \}$,
and
$\DataSet^{(i)} (\zeit) =
\{\svecproj (\zeit),
    \svecproj (\zeit + t_s),
    \ldots,
    \svecproj (\zeit + (N_i - 1)t_s)
\}$
be states sampled at rate $t_s^{-1}$
along one period of
$\po_i$ and a chaotic trajectory beginning at time $\zeit$,
respectively, and
$\hat{}$\ indicate the projection onto the bases of $\po_i$.
We define the shadowing distance $\distShadow^{(i)} (\zeit)$
of turbulence from
$\po_i$ at time $\zeit$ as
\begin{equation}
    \distShadow^{(i)} (\zeit) = \sum_{k=0}^{1}
    w_k W_{\infty} 
    \left( \PerD^{(i)}_k (\zeit), \PerD^{(\po_i)}_k \right)
    \label{e-distShadow}
\end{equation}
where $\PerD^{(\po_i)}$ and $\PerD^{(i)} (\zeit)$ are the persistence
diagrams obtained from
$\DataSet^{\po_i}$ and
$\DataSet^{(i)} (\zeit)$, respectively, and $w_{0,1}$ are the
weights of respective contributions from the components and holes.
In
what follows, these weights are set to
\(
    w_{0,1} = [W_{\infty} ({\rm D}, \PerD^{(\po_i)}_0)
    + W_{\infty} ({\rm D}, \PerD^{(\po_i)}_1)]^{-1} \,, \label{e-weights}
\)
where ${\rm D}$ denotes the empty persistence diagram with
diagonal elements only.
This choice of the weights sets the shadowing
distance of a 
\textpo{} to an empty data set
to $1$; thus renders the shadowing distances from different 
\textpo{}s comparable.
As an illustration, \reffig{f-ShadowDecomp}(a) shows the shadowing
distances of a turbulent trajectory
from $8$ out of $17$
\textpo{}s.

We expect the local minima of $S^{i} (\zeit)$ to correspond to the
episodes of turbulent flow shadowing $\po_i$.
Following this assumption, we define the
\emph{shadowing decomposition} of a turbulent flow in a time
interval $\zeit \in [\zeit_0, \zeit_f]$ over 
$\{\po_1, \po_2, \ldots \po_{N_{\po}} \}$
for a threshold distance $S_{{\rm th}}$
by the following algorithm.
Starting at time $\zeit = \zeit_0$,
we find $i_{\min} = \argmin_i S_i (\zeit)$.
If $\distShadow_{i_{\min}} (\zeit)$ is less than
$\distShadow_{th}$, then we save the pair
$(t, i_{\min})$ and increase $\zeit$
by 
$T_{i_{\min}}$;
otherwise, we increase $\zeit$ by $\zeit_s$
and repeat the procedure until the final time $\zeit_f$ is reached.
The result 
is the set of pairs
$(\zeit, i_{\min})$ which we 
interpret as ``turbulence at time interval $[t, t+T_{i_{\min}}]$ 
can be approximated by $\po_{i_{\min}}$.''
In \reffig{f-ShadowDecomp}(b), we visualized the shadowing decomposition 
($\distShadow_{\rm th} = 0.5$) of turbulence corresponding to the 
same episode as \reffig{f-ShadowDecomp}(a)  as a bar plot where the 
length of each bar is equal to the period of the respective \textpo{}.
Supplementary video~\footnote{See Supplemental Material at 
    [URL will be inserted by publisher] for details on 
    the 
    (i) adequacy of our resolution, 
    (ii) effective number of degrees of freedom,
    (iii) \textpo{}s and their selection for modeling, 
    (iv) demonstration of partial shadowing of a \textpo{}, 
    (v) robustness against threshold choice, 
    and (vi) convergence of statistics and models.}
shows another visualization of this decomposition
for $\zeit \in [0, 100]$
where velocity and vorticity isosurfaces of turbulence are shown next to those
of the 
\textpo{}s that are being shadowed along with their state space
projections.
As can be seen in the supplementary 
video (also demonstrated in SM\rf{Note2}), 
our decomposition is able to 
generate shadowing signals even when  turbulence follows a
\textpo{} for less than a full period.

From its shadowing decomposition, 
we can infer a model
of the turbulent flow as a Markov chain\rf{Papoulis2002} with
the transition matrix $\TransMat$, whose elements
$\TransMat_{ij}$ correspond to the probability of
shadowing $\po_j$ after $\po_i$.
We estimated these probabilities from 
$18$ different runs, separate from those used to find 
the
\textpo{}s, with a total run time of
$\zeit_{\rm tot} = 25039$ excluding the initial transients and the 
laminarization events\rf{Note2}.
For the threshold choice 
$S_{\rm th} = 0.5$, we found the shadowing
events to cover $75 \%$ of the total time.
As a robustness test, we repeated our computations for
$S_{\rm th} \in [0.4, 0.6]$.
While the fraction of turbulent time that is covered by the
\textpo{}s differs for different 
$S_{\rm th}$,
it remains always above $50 \%$ 
($54\%$ for $S_{\rm th} = 0.4$ and $88\%$ for $S_{\rm th} = 0.6$) and the
transition probabilities of the Markov process vary only slightly\rf{Note2}.
Therefore, our results in the following are not sensitive to
this threshold. 

\begin{figure}[h]
    \begin{overpic}[height=0.29\linewidth]{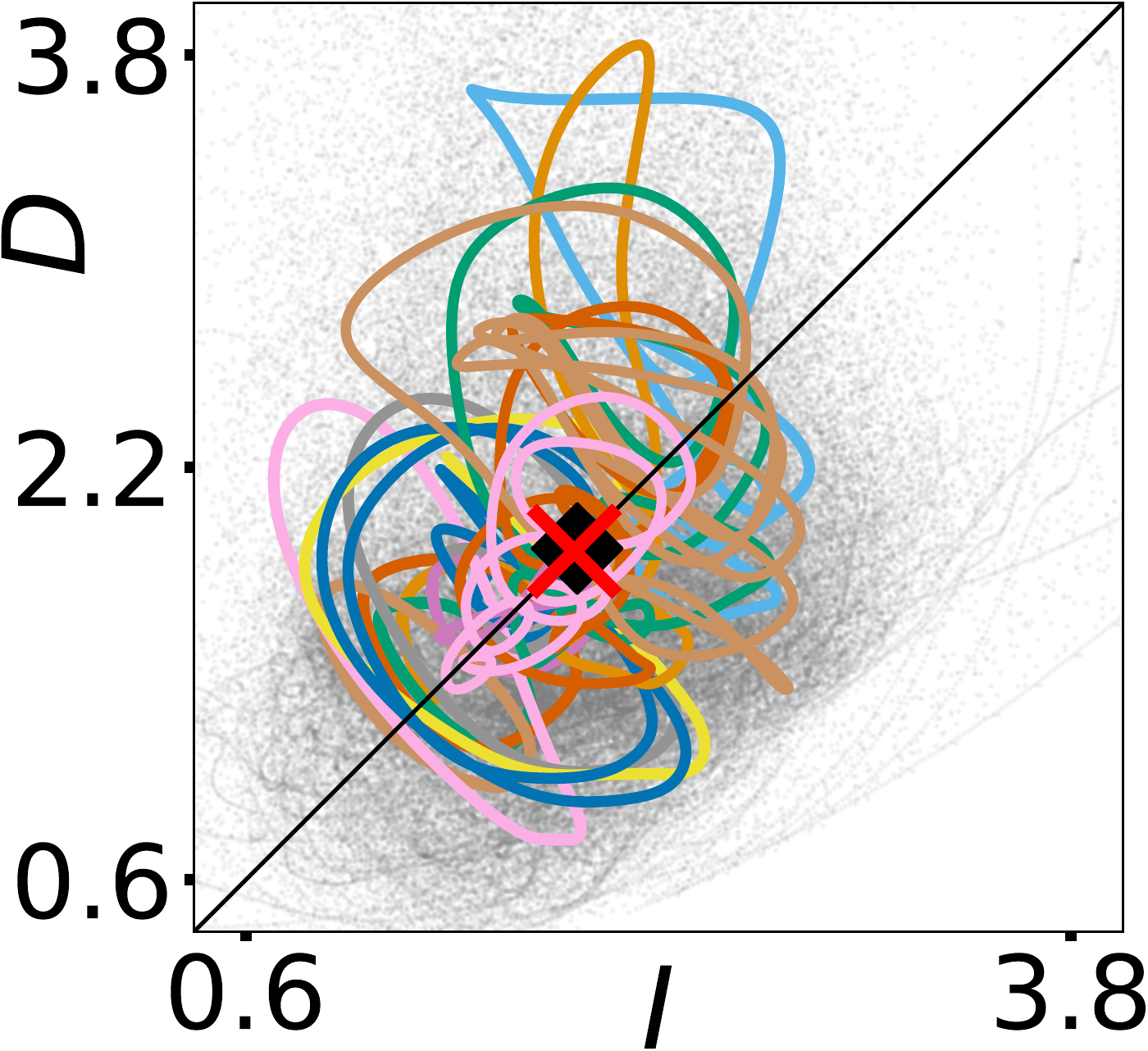}
        \put (0,0) {(a)}
    \end{overpic}
    \begin{overpic}[height=0.29\linewidth]{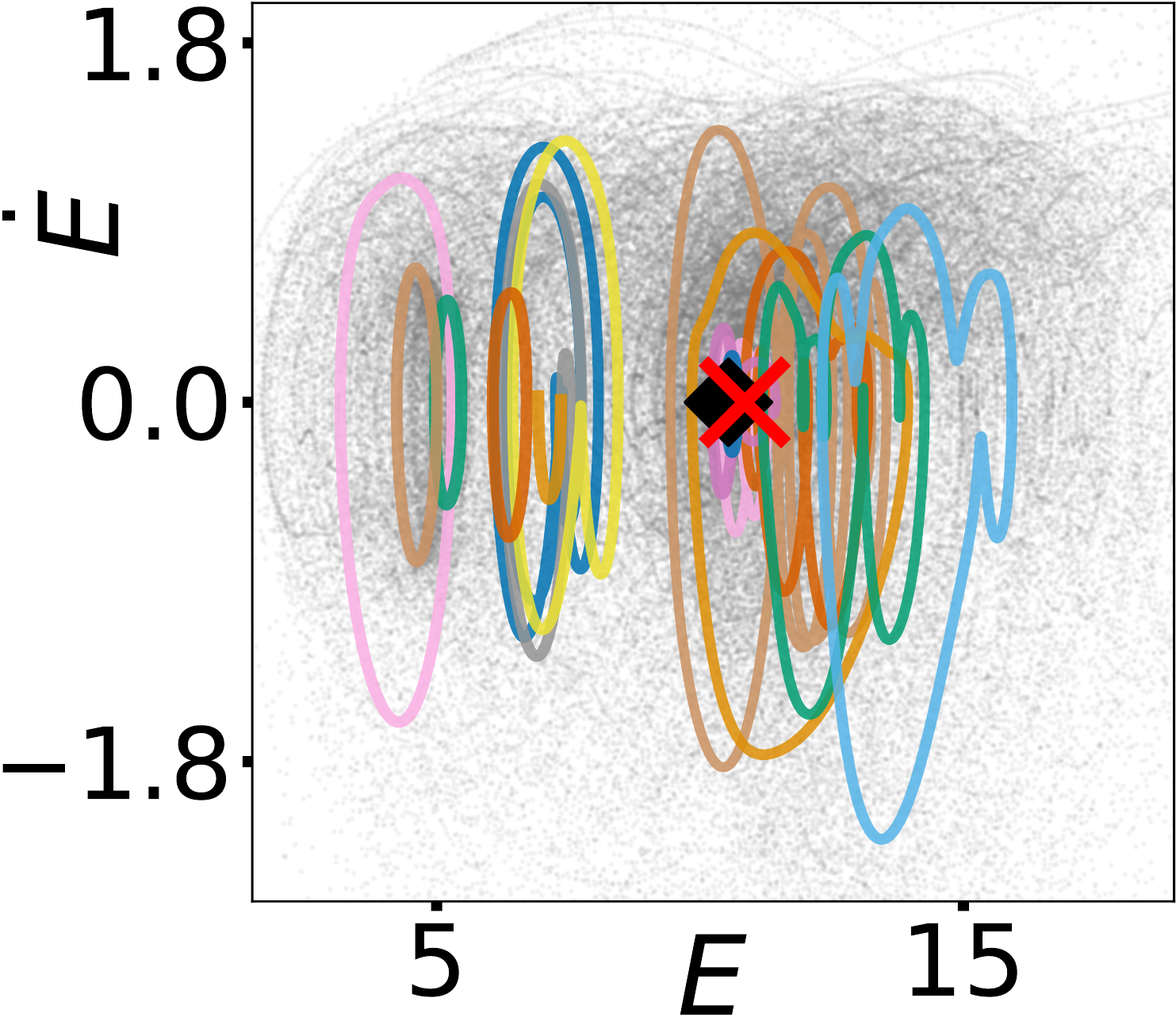}
        \put (0,0) {(b)}
    \end{overpic}
    \begin{overpic}[height=0.295\linewidth]{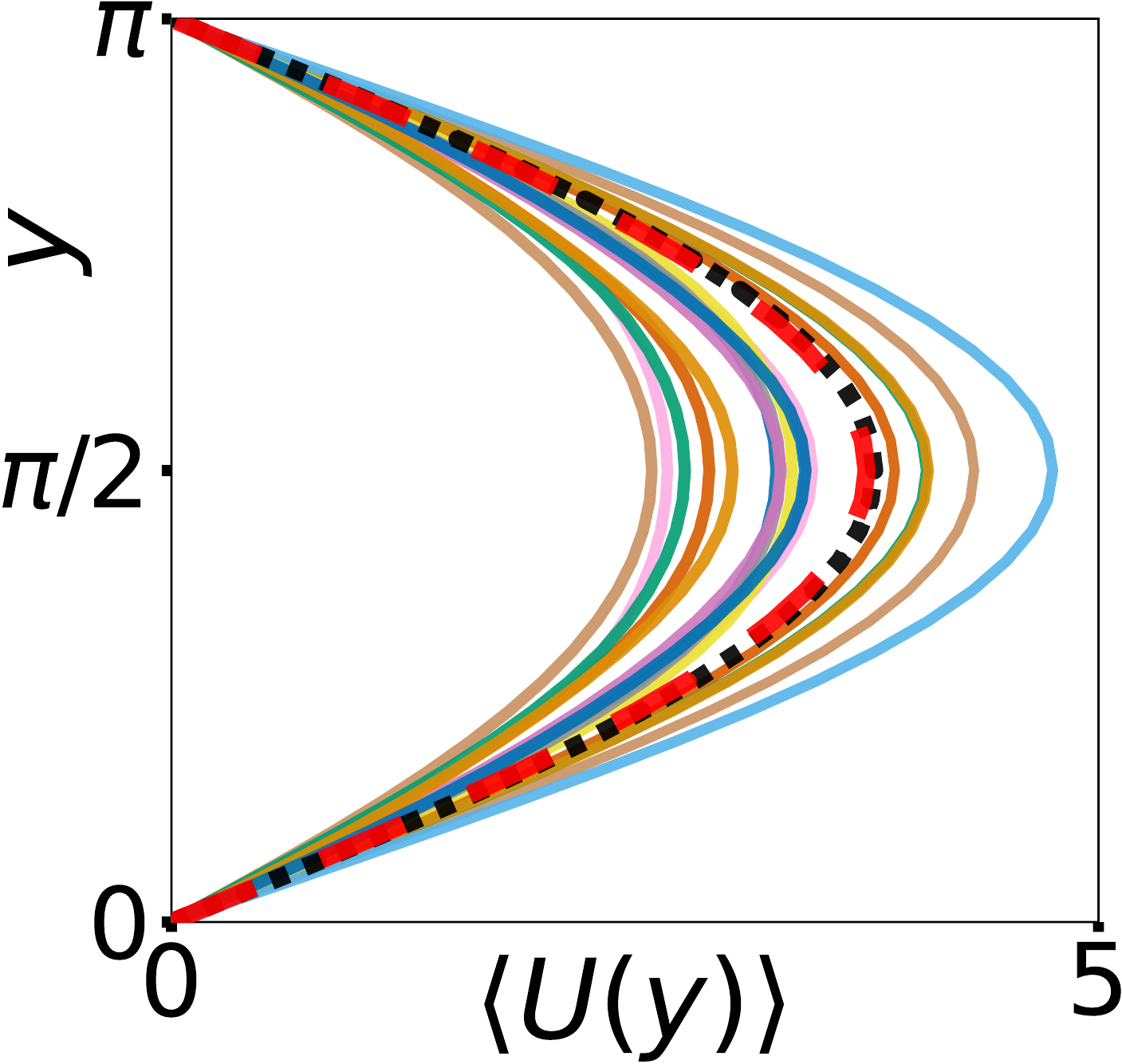}
        \put (0,0) {(c)}
    \end{overpic}
    \caption{(Color online)
        Turbulent trajectories (dots, gray) and
        \textpo{}s (loops, colors) visualized
        (a) on the $ID$
        and (b) the $E\dot{E}$ planes 
        along with the long-time averages 
        $\langle D \rangle_{\infty} = \langle I \rangle_{\infty} = 1.885$,
        $\langle E \rangle_{\infty} = 10.54$, 
        and $\langle \dot{E} \rangle_{\infty} = 0$ 
        (diamonds, black) 
        and the PO estimates 
        $\langle D \rangle_{\InvDist} = \langle I \rangle_{\InvDist} = 1.874$,
        $\langle E \rangle_{\InvDist} = 10.85$, and $\langle \dot{E} \rangle_{\InvDist} = 0$
        (crosses, red).
        (c) Velocity profiles
        averaged over POs
        (solid, colors) along with the 
        long-time average $\langle U (y) \rangle_{\infty}$ (dotted, black)
        and its PO estimate 
        $\langle U (y) \rangle_{\pi}$ (dashed, red).
        Only the half domain $y \in [0, \pi]$ is shown in (c) 
        since the other half corresponds to its mirror image.
    \label{f-Global}}
\end{figure}

The invariant distribution $\InvDist$ of the Markov chain is the
left eigenvector of $\TransMat$ with unit eigenvalue, satisfying
the normalization condition $\sum_i \InvDist_i = 1$.
\reffig{f-ShadowDecomp}(c) shows a network visualization of the
Markov chain that models the $3D$ Kolmogorov flow
that we studied here.
Each node of \reffig{f-ShadowDecomp}(c) corresponds to a
\textpo{} 
with the size of the node $i$ proportional to
$\InvDist_i$ and the directed edges indicate possible transitions
from one
\textpo{} to the next.
The nodes have also self-loops (not shown in \reffig{f-ShadowDecomp}(c) for 
clarity) that correspond to close recurrence events in which turbulence shadows 
a \textpo{} for more than one period.
From
$\InvDist$, we can predict
long-time averages of the turbulent flow's observables in terms of
their values computed over
\textpo{}s.
Let $\Observable$ be an observable
and $\langle \Observable \rangle_i $ be its average
over  $\po_i$ and its symmetry copies.
The long-time average $\langle \Observable \rangle_\infty$ can be approximated
as
\begin{equation}
    \langle \Observable \rangle_\InvDist =
        \frac{\sum_{i = 1}^{N_{\po}} \InvDist_i T_i \langle \Observable \rangle_i}{
                \sum_{i = 1}^{N_{\po}} \InvDist_i T_i } \,,
    \label{e-MeanObservable}
\end{equation}
where we interpret the coefficients $\pi_i T_i$ as the mean time that
chaotic flow spends in the neighborhood of  $\po_i$.
The observables
that we consider are
kinetic energy
    $E = \| \vfield \|^2 / 2$,
power input
    $I = \frac{1}{L_x L_y L_z} \int \vfield \cdot \mathbf{f}\ d\mathbf{x}$,
dissipation
    $D = \nu \|\nabla \times \vfield \|^2 $,
and the velocity profile $U(y) = \frac{1}{L_x L_z} \int \int u(x, y, z)\ dx dz $.
\reffig{f-Global}(a,b) show the \textpo{}s and the data
sampled from turbulence on $ID$ and $E\dot{E}$ planes ($\dot{E} = I - D$)
respectively. 
The long-time averages
$\langle D \rangle_{\infty} = \langle I \rangle_{\infty} = 1.885$,
$\langle E \rangle_{\infty} = 10.54$, and $\langle \dot{E} \rangle_{\infty} = 0$ 
along with the \textpo{} estimates \refeq{e-MeanObservable}
$\langle D \rangle_{\InvDist} = \langle I \rangle_{\InvDist} = 1.874$,
$\langle E \rangle_{\InvDist} = 10.85$, and $\langle \dot{E} \rangle_{\InvDist} = 0$
are also marked in \reffig{f-Global}(a,b).
In \reffig{f-Global}(c), we plot the mean \textpo{} velocity profiles 
along with the long-time 
average $\langle U(y) \rangle_\infty$ and its PO estimate
$\langle U(y) \rangle_{\InvDist}$ \refeq{e-MeanObservable}.
These long-time ($t_{\rm total} = 35492$)
averages are computed over runs that are separate from those used 
to infer the transition probabilities.

As noted above and shown in \reffig{f-Global}, the \textpo{} 
estimates \refeq{e-MeanObservable} of observables agree 
(to 2 digits for $E$, $I$, and $D$)
with the long-time averages, providing an 
a posteriori verification of our reduced-order model.
It is worth emphasizing
that  
the observable
averages over individual POs
can be quite different (\reffig{f-Global}, see also \rf{Note2})
from the long-time averages. Thus, it is
crucial for the weights in the 
sum \refeq{e-MeanObservable} to be correct for numerical agreement.
Interestingly, 
we found the orbits with long periods to 
be necessary 
to capture the 
long-time averages since our shortest $9$
\textpo{}s 
with periods less than $10$ have mean dissipation 
rates less than the long-time average  
$\langle D \rangle_{\infty} = 1.885$\rf{Note2}. 
This observation is at odds with our
intuition based on the cycle expansions of strange sets\rf{PC88} 
where the long
\textpo{}s appear only in correction
terms.

One feature of the $3D$ Kolmogorov flow that we
do not capture in our model is the laminarization events
since we exclude them from our training data.
Therefore, the Markov chain and its invariant distribution
should be understood as the
model of the nonattracting chaotic set\rf{LaiTel2011}
underlying transient turbulence and the natural measure
over it, respectively.
We note that the consistency of the long-time
averages with those computed
using \refeq{e-MeanObservable} is evidence of
ergodicity for this chaotic set.

In this study, we combined ideas from
the dynamical systems theory with topological data analysis to produce a
low-dimensional turbulence model, wherein the dynamics is 
viewed as a Markov chain of shadowing events.
We confirmed the accuracy of this description by reproducing 
the long-time averages of the flow's observables from the 
invariant distribution of the Markov process.
We would like to note 
that coarse-grained models such as ours can be utilized in
control
methods that drive the system towards desired state space regions.
In conclusion, we believe that modeling turbulence using
\textpo{}s
not only deepens our understanding of it but also opens new avenues for
applications.

\begin{acknowledgments}
We thank the referees for improving this paper with their 
comments.
We acknowledge stimulating discussions with H.\ Edelsbrunner.
This work was supported by a grant from the Simons Foundation 
(662960, BH).
The numerical calculations were performed at
TUBITAK ULAKBIM High Performance and Grid Computing Center (TRUBA resources)
and IST Austria High Performance Computing cluster.
\end{acknowledgments}

\bibliography{shadow}

\onecolumngrid
\newpage

\widetext
\begin{center}
	\textbf{\large Supplemental Material}
\end{center}
\setcounter{equation}{0}
\setcounter{figure}{0}
\setcounter{table}{0}

\renewcommand{\theequation}{S\arabic{equation}}
\renewcommand{\thefigure}{S\arabic{figure}}
\renewcommand{\thetable}{S\Roman{table}}

\input{shadow_supplemental_content}

\end{document}

%% file: setup.tex
\usepackage[T1]{fontenc}
\usepackage[utf8]{inputenc}
\usepackage[english]{babel}
\addto\captionsenglish{}
\addto\captionsenglish{}
\usepackage{graphicx,amssymb,amsbsy}
\usepackage[percent]{overpic}
\usepackage[colorlinks]{hyperref}
\usepackage{amsmath}
\usepackage[section]{placeins} 
\usepackage{siunitx} 

\graphicspath{{figs/}}  

%% file: def.tex
\newcommand{\rf}     [1] {~\cite{#1}}
\newcommand{\refref} [1] {Ref.~\cite{#1}}

\newcommand{\refrefs}[1] {Refs.~\cite{#1}}

\newcommand{\reffig} [1] {Fig.~\ref{#1}}

\newcommand{\reftab} [1] {Table~\ref{#1}}

\newcommand{\refeq}  [1] {(\ref{#1})}

\newcommand{\etal}{{\em et al.}}    
\newcommand{\ie}{{i.e.}}        

\newcommand{\vfield}{\ensuremath{\mathbf{u} }}
\newcommand{\Fvfield}{\ensuremath{\hat{\mathbf{u}}}}

\newcommand{\zeit}{\ensuremath{t}}  
\newcommand{\Reynolds}{\textit{Re}}  

\newcommand{\inprod}[2]{\left\langle #1 ,\, #2 \right\rangle}
\newcommand{\svec}{\ensuremath{\xi}}
\newcommand{\svecproj}{\ensuremath{\hat{\xi}}}
\newcommand{\po}{\ensuremath{\overline{{\rm po}}}}

\newcommand{\identity}{\ensuremath{I}}
\newcommand{\resol}{\ensuremath{r}}
\newcommand{\DataSet}{\ensuremath{\Xi}}
\newcommand{\PerD}{\ensuremath{{\rm PD}}}
\newcommand{\distShadow}{\ensuremath{S}}
\newcommand{\TransMat}{\ensuremath{P}}
\newcommand{\InvDist}{\ensuremath{\pi}}
\newcommand{\Observable}{\ensuremath{\Omega}}

\DeclareMathOperator*{\argmin}{arg\,min}

\newcommand{\continue}{\nonumber \\ }
\newcommand{\bea}{\begin{eqnarray}}
\newcommand{\eea}{\end{eqnarray}}
\newcommand{\beq}{\begin{equation}}
\newcommand{\eeq}{\end{equation}}

%% file: shadow_supplemental_content.tex
\section{Numerical representation and the adequacy of resolution}

\begin{figure}[h]
    (a)\includegraphics[width=0.3\linewidth]{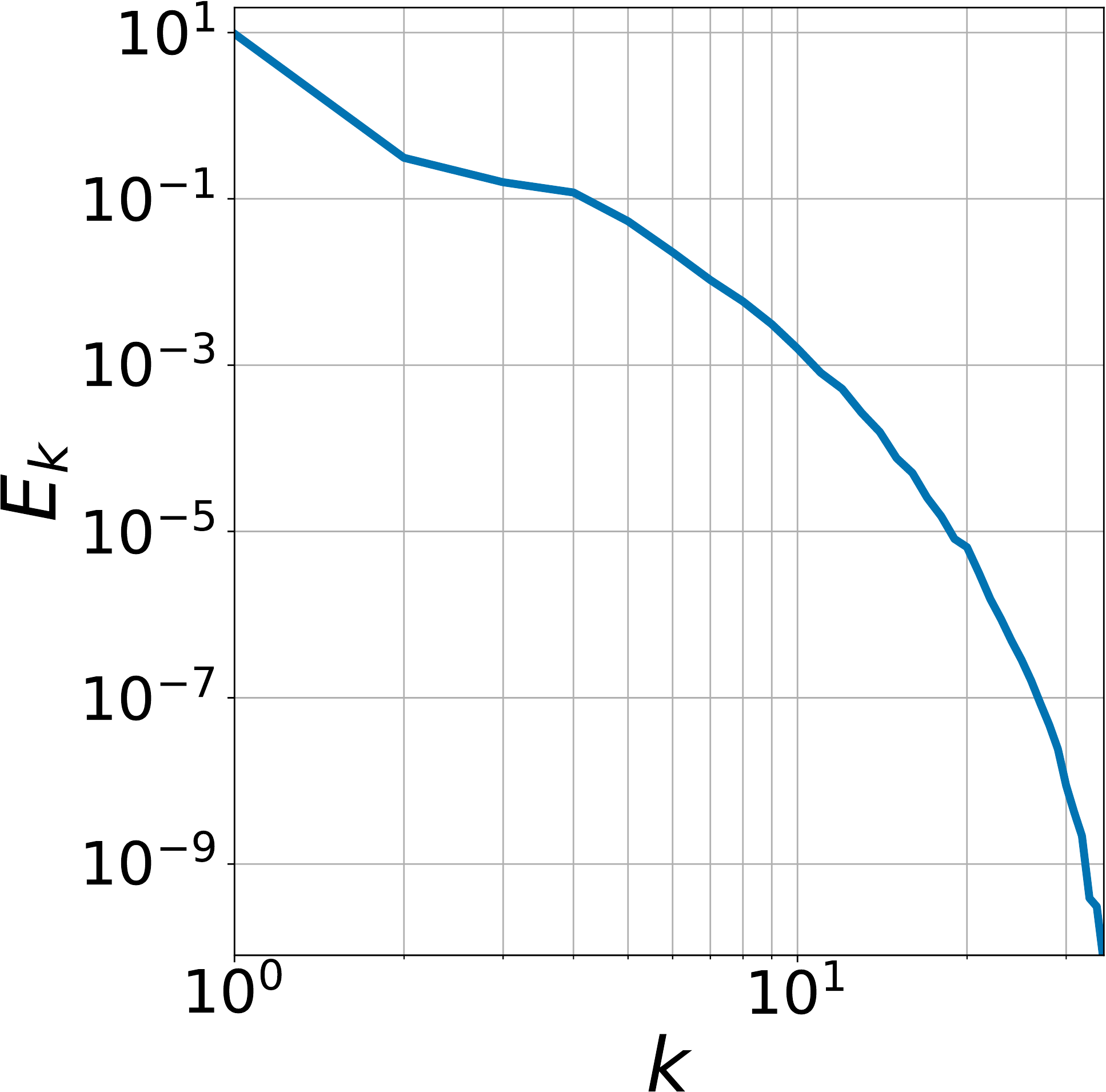}
	(b)\includegraphics[width=0.3\linewidth]{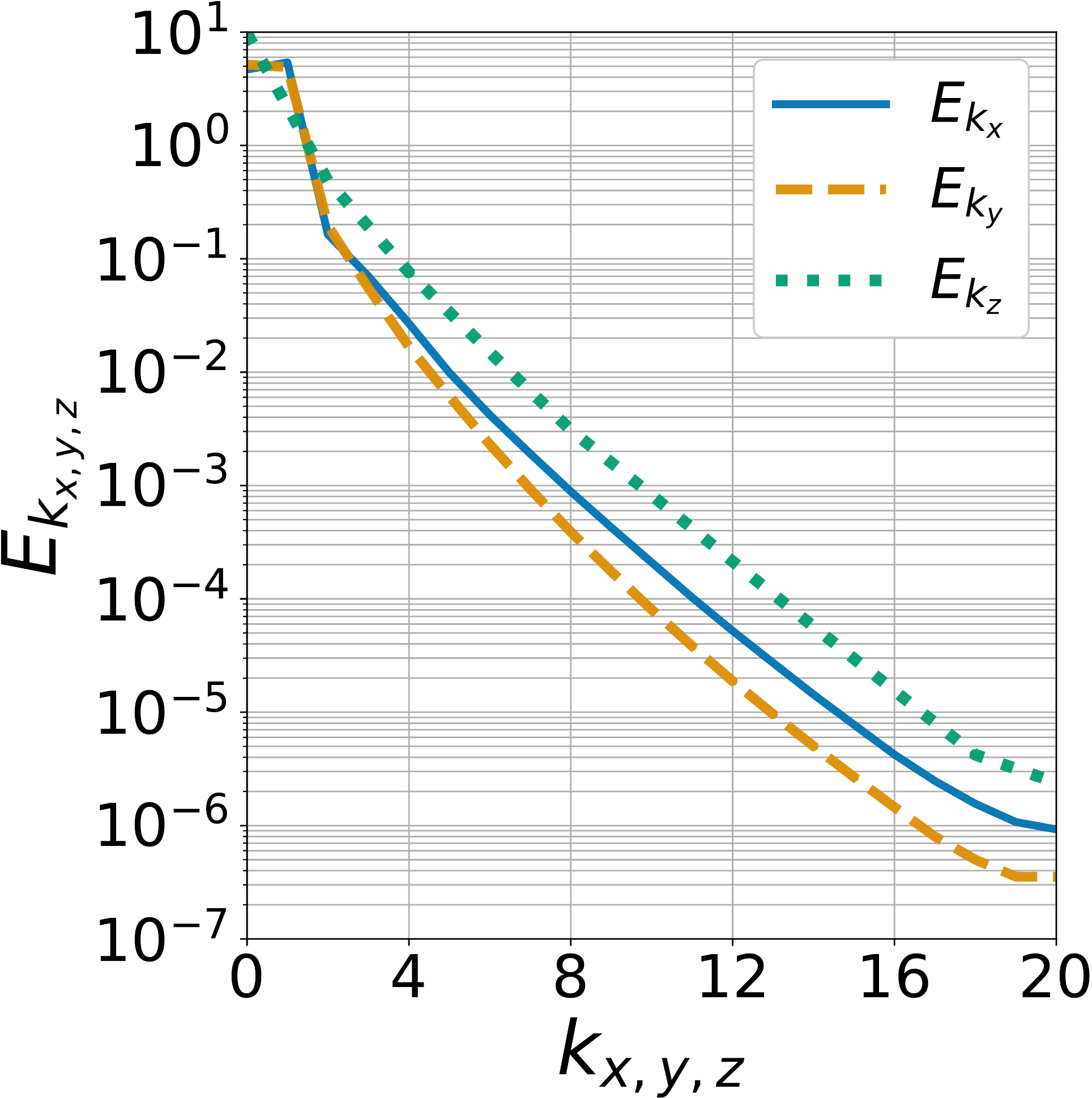}
	(c)\includegraphics[width=0.3\linewidth]{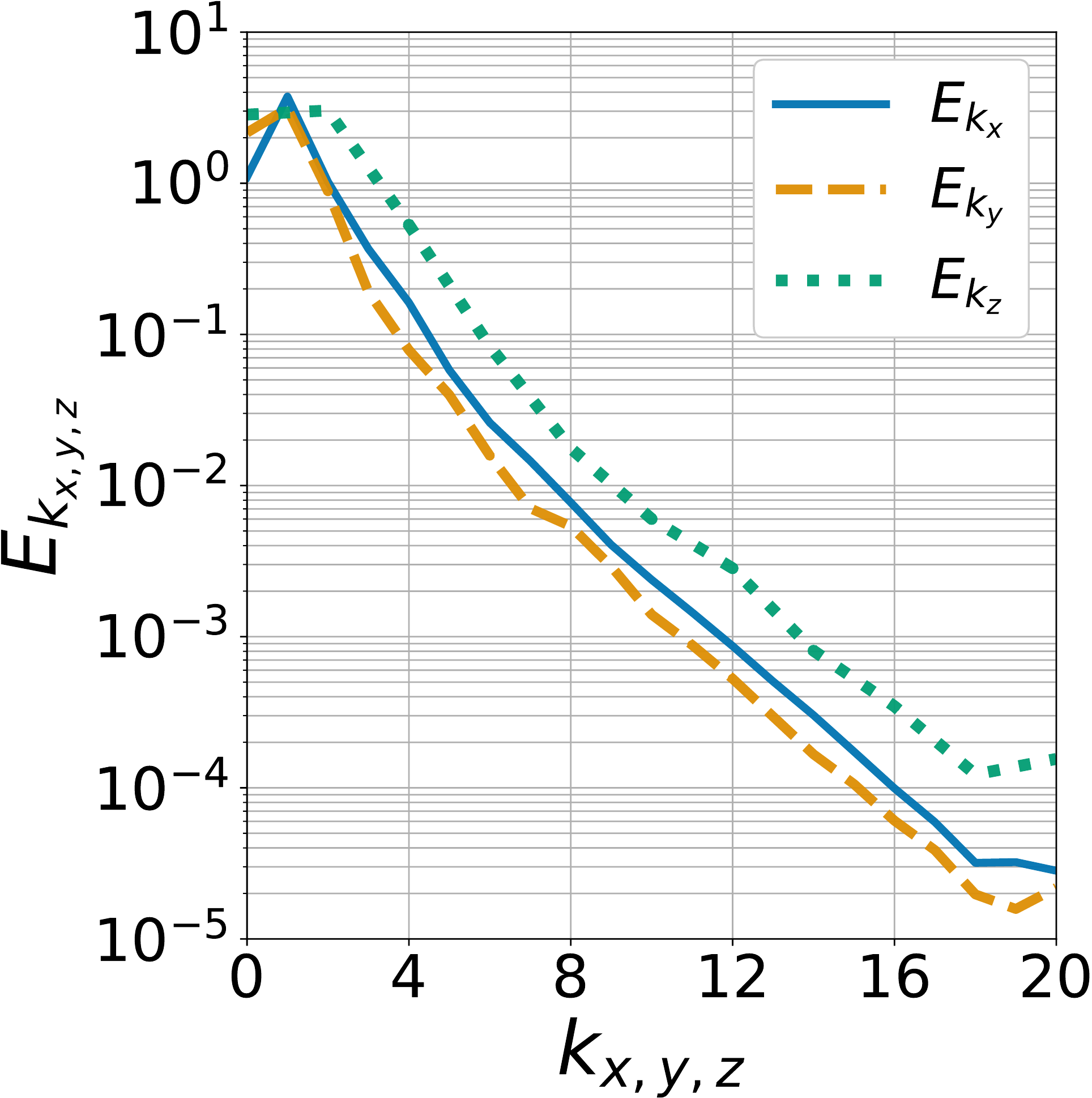}
	\caption{
		(a) Time- and shell-averaged energy spectrum. 
		(b) Time-averaged energy spectra in each spatial direction.
		The time averages are obtained by averaging the spectrum of 
		$3012$ distinct turbulent state sampled every 
        $\Delta t = 40$ from 21 runs.
        (c) Example of a ``worst-case'' energy spectra where the 
        spectral drop-off in $x$ direction is $4$ orders of magnitude.
		\label{f-spectra}}
\end{figure}

We represent the velocity fields as a Fourier expansion
\beq
	\vfield(x, y, z) = \sum_{k_x, k_y, k_z}
					   \Fvfield[k_x, k_y, k_z] e^{i (k_x x + k_y y + k_z z)} \,, 
\eeq
where the sum is carried through all resolved wave numbers. 
In our $[2\pi \times 2\pi \times \pi]$ domain with 
$[64 \times 64 \times 32]$ grid points, 
after the $2/3$ dealiasing, the nonzero wave numbers in $x$, $y$, and 
$z$ directions are
\begin{equation}
\begin{aligned}
k_x &= \{-20, -19, \ldots, 19, 20\}\,,\\
k_y &= \{-20, -19, \ldots, 19, 20\}\,,\\
k_z &= \{-20, -18, \ldots, 18, 20\}\,.
\end{aligned}
\end{equation}
In the literature, the energy spectrum of turbulent flows is usually 
presented in the time- and shell-averaged form as shown in
\reffig{f-spectra}(a), where $E_k$ is 
the total energy contained in Fourier coefficients with amplitudes falling in the 
interval $[k, k+1)$. 
Note that the wave numbers shown in \reffig{f-spectra}(a) go beyond 
the largest resolved wave number $k_{max}=20$ in each direction since the 
amplitude of the wave vector $[k_x = 20, k_y = 20, k_z = 20]$ is 
$|\mathbf{k}|_{max} \approx 34.64$.
In order to demonstrate the adequacy of our resolution, we therefore show 
the time-averaged spectra  individually for each spatial direction 
in \reffig{f-spectra}(b), 
where one can see at least $6$ orders of magnitude drop-off
from the longest resolved wave length to the shortest one. 
Even though the time-averaged spectra of \reffig{f-spectra} show
$6$ orders of magnitude or more drop-off, 
we observed that this spectral gap can be as low as $4$ orders 
instantaneously for states with high dissipation.
We show the instantaneous energy spectra for one such state in 
\reffig{f-spectra}(c).

As a further test of the adequacy of our numerical representation, 
we progressively increased the spatial and temporal resolution up to 
a factor of $3/2$ and confirmed that our periodic orbits continue to 
exist at these higher resolutions by Newton-converging them at every 
step. 
In our preliminary work where we used a lower spatial resolution 
$[N_x = 32, N_y = 32, N_z = 16]$, some of the periodic orbits that 
we found did not pass this test at various intermediate steps 
leading us to our final spatial resolution $[64,64,32]$. 

Our spatial resolution is slightly lower than that of 
\refref{vanVeenGoto2016}, where the authors reported their results 
using the Reynolds number 
$\Reynolds = L_y \sqrt{L_y \gamma} / 2 \pi \nu$ 
as the control parameter. 
Following this definition, our choice of parameters yields 
$\Reynolds \approx 50$, which is much lower than 
$\Reynolds = 170$ at which \refref{vanVeenGoto2016} 
reports long-lived turbulent transients.  
At first glance, our observation of long-lived turbulent transients 
at $\Reynolds \approx 50$ might seem to contradict with the results of \refref{vanVeenGoto2016}. 
However, the possibility of observing turbulence in this system 
at \Reynolds{} as low 
as $\Reynolds = 40$ was already reported in the discussion of 
\refref{vanVeenGoto2016}, where the authors speculated the existence 
of edge states other than the one they reported. 
We confirmed their observation by matching our domain exactly 
to theirs by extending the spatial extent in the $z$ direction to 
$L_z = 2\pi$ and lifting the symmetry constraints that we impose on the 
dynamics. 
Our numerical experiments indicate that it is possible to observe 
long-lived turbulent transients with life times of $O(1000)$ in such 
domains. 

\section{Effective number of degrees of freedom}
The wave numbers that are kept in our numerical integrator at 
\(L_x \times L_y \times L_z = 2 \pi \times 2 \pi \times \pi\),
excluding dealiased modes, are
\begin{equation}
\begin{aligned}
k_x &= \{-20, -19, \ldots, 19, 20\}\,,\\
k_y &= \{-20, -19, \ldots, 19, 20\}\,,\\
k_z &= \{0, 2, \ldots, 18, 20\}\,,
\end{aligned}
\end{equation}
where we omit the $k_z < 0$ part of the spectrum since they can be 
recovered from the condition $\Fvfield^*[-\mathbf{k}] = \Fvfield[\mathbf{k}]$
the expansion coefficients obey as the velocity field in the physical space is real valued.
This gives \(41\times41\times11\times2\times3=110946\) numerical 
degrees of freedom, where the factors of $2$ and $3$ correspond to 
real and imaginary parts of the Fourier coefficients and the dimensions 
of physical space, respectively. 
Although this discretization implies a
$110946$-dimensional dynamical system, in practice, the number of 
degrees of freedom in our system is effectively reduced by the
divergence-free 
condition and the imposed symmetries.
Since the $\mathbf{k} = 0$ mode is time-invariant (Galilean invariance)
we set it to $0$, which eliminates $6$ degrees of freedom from the Fourier series.
The divergence-free condition $i \mathbf{k} \cdot \Fvfield = 0$ reduces the 
independent number of degrees of freedom by a factor of $2/3$, since the 
knowledge of two velocity field components uniquely determines the third one. 
In addition, restricting the dynamics into a subspace that is invariant under the symmetries
$S_x$ and $R_z$ further drop the degrees of freedom by a factor 
of $4$, resulting in the final number of independent 
degrees of freedom $[(110946 - 6) \times 2 / 3] / 4 = 18490$.

The number of independent numerical degrees of freedom is still much higher 
than the manifold in which the turbulent dynamics takes place. 
Although, we do not have a rigorous proof of existence of such an inertial 
manifold, we think that it is reasonable to assume its existence due to the 
dissipation in the system.
While estimating the dimension of this manifold is beyond the scope of the 
present work, we here provide an evaluation of various lower-dimensional 
embeddings based on principal component analysis. 
To this end, we first compute $24096$ principal components corresponding 
to $3012$ uncorrelated turbulent states and their $8$ discrete symmetry copies. 
We then construct embeddings of different dimensions using the leading $d$ principal
components and compute the error 
\beq
	\epsilon_{\rm PCA} = \|\vfield - {\mathbf P} \vfield \| / \| \vfield \| \,, \label{e-PCAerr}
\eeq
where $\vfield$ is sampled from a test set distinct from those that were used 
to construct the principal components and ${\mathbf P}$ denotes the 
projection onto the principal components.
Table \ref{t-bases-nosymred-captures} shows the minimum, maximum, and 
mean errors on this test set for different embedding dimensions. 

\begin{table}[h]
	\centering
	\caption{
		Minimum, maximum, and mean error \refeq{e-PCAerr} for different 
		embedding dimensions computed for a test turbulent trajectory with 
		a lifetime $t = 1924.0$, sampled at $t_s = 0.1$. 
		\label{t-bases-nosymred-captures}
	}
	\begin{ruledtabular}
		\begin{tabular}{c c c c}
			\(d\)   &   $\min \epsilon_{\rm PCA}$ & $\max \epsilon_{\rm PCA}$ & $\langle \epsilon_{\rm PCA} \rangle$ \\
			\hline
			\(64\)   &   \(1.224\times10^{-2}\) &   \(0.3884\) &   \(0.07880\)  \\
			\(128\)  &   \(6.006\times10^{-3}\) &   \(0.2775\) &   \(0.05151\)  \\
			\(256\)  &   \(2.866\times10^{-3}\) &   \(0.1774\) &   \(0.03101\)  \\
            \(512\) &   \(1.150\times10^{-3}\) &   \(0.1188\) &   \(0.01696\) \\
			\(1024\)  &   \(4.468\times10^{-4}\) &   \(0.07208\) &   \(0.008266\)  \\
			\(2048\) &   \(1.593\times10^{-4}\) &   \(0.03930\) &   \(0.003540\)  \\
            \(4096\) &   \(4.393\times10^{-5}\) &   \(0.02005\) &   \(0.001268\)
		\end{tabular}
	\end{ruledtabular}
\end{table}
\FloatBarrier

\section{Selection of the periodic orbits}

In \reftab{tab:list-periodicCompact} we list the 
period,
mean kinetic energy and dissipation, 
and the contribution to the invariant distribution of
the 18 periodic orbits that we found from near recurrences of 
the turbulent flow. 
We began our modeling trials using the subset of periodic orbits
with periods shorter than $10$, however, all of these attempts 
resulted in estimates of dissipation 
lower than its long-time average. 
The reason behind this is readily seen in
\reftab{tab:list-periodicCompact} where all of the periodic 
orbits with $T<10$ have mean rate of dissipation less than 
the long-time average $\langle D \rangle_{\infty} = 1.885$.
Consequently, we decided to use all numerically-found periodic orbits.
However, as we shall explain in the following, we found that excluding 
$\po_{18}$ from the model did not result in a significant 
change in our final results. 

\begin{table}[h]
	\centering
	\caption{List of periodic orbits.
		Shown are the period \(T\),
		average kinetic energy \(\langle E \rangle\),
        and average dissipation \(\langle D \rangle\), 
        and contribution to the invariant distribution \(\pi\).
		\label{tab:list-periodicCompact}
	}
	\begin{ruledtabular}
        \begin{tabular}{ccccc}
            \multicolumn{1}{c}{i} 
          & \multicolumn{1}{c}{\(T\)} 
          & \multicolumn{1}{c}{\(\langle E \rangle\)} 
          & \multicolumn{1}{c}{\(\langle D \rangle\)} 
          & \multicolumn{1}{c}{\(\pi\)}\\
            \hline
            \input{list-periodic-orbits}
        \end{tabular}
	\end{ruledtabular}
\end{table}
\FloatBarrier

\begin{figure}[h]
    \begin{overpic}[height=0.235\linewidth]{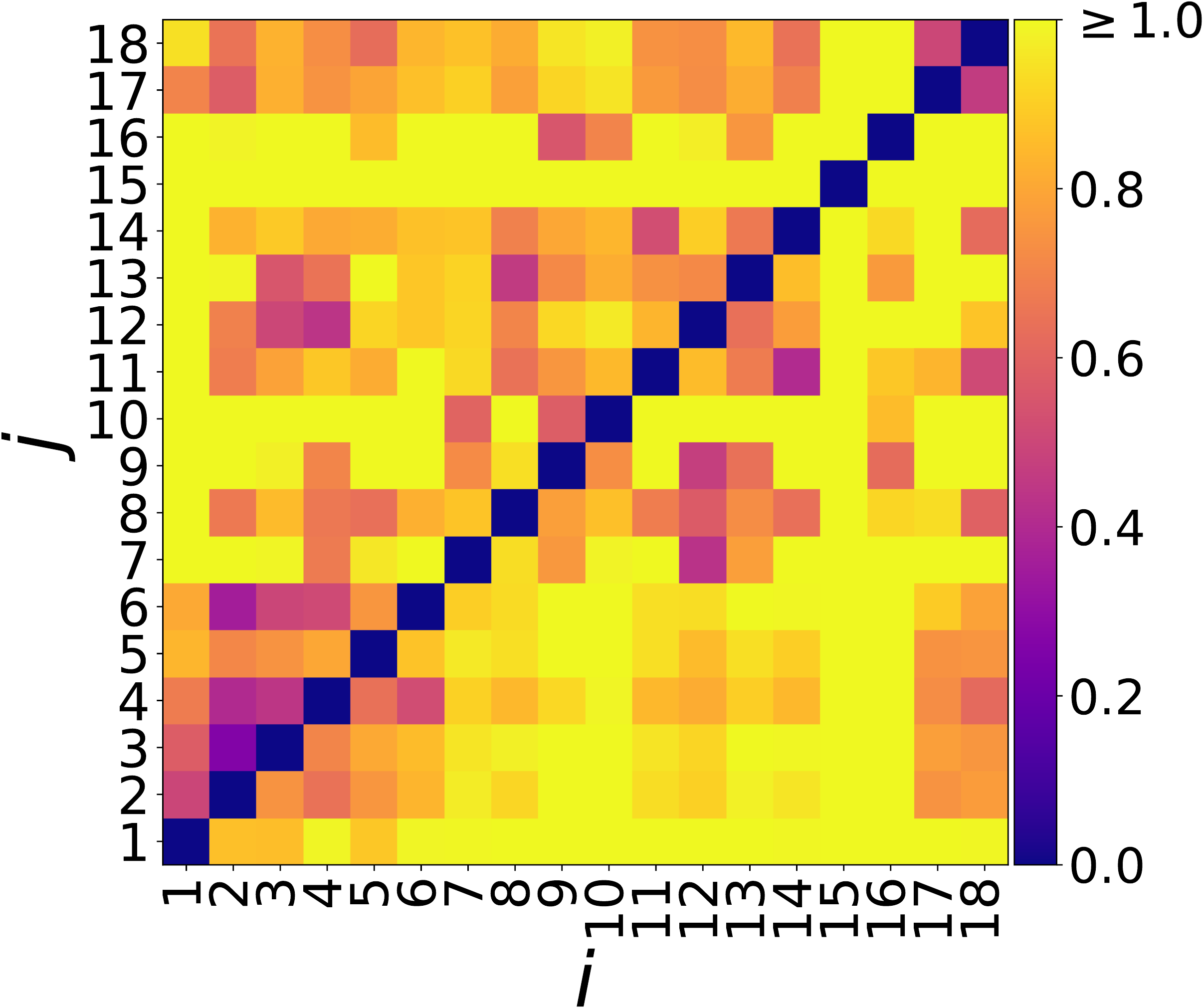}
    \put (0,0) {(a)}
    \end{overpic}
    \begin{overpic}[height=0.235\linewidth]{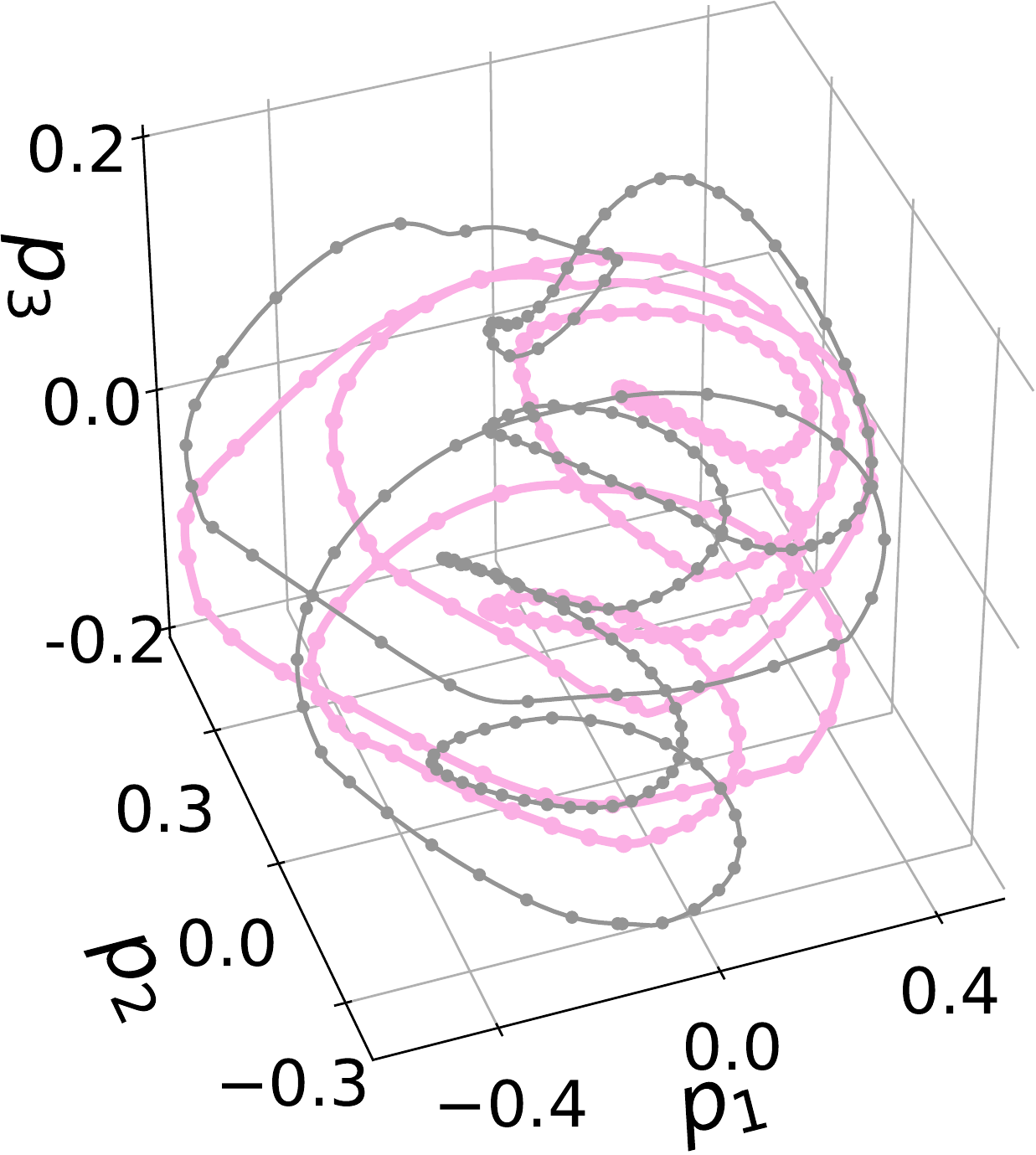}
        \put (0,0) {(b)}
    \end{overpic}
    \begin{overpic}[height=0.235\linewidth]{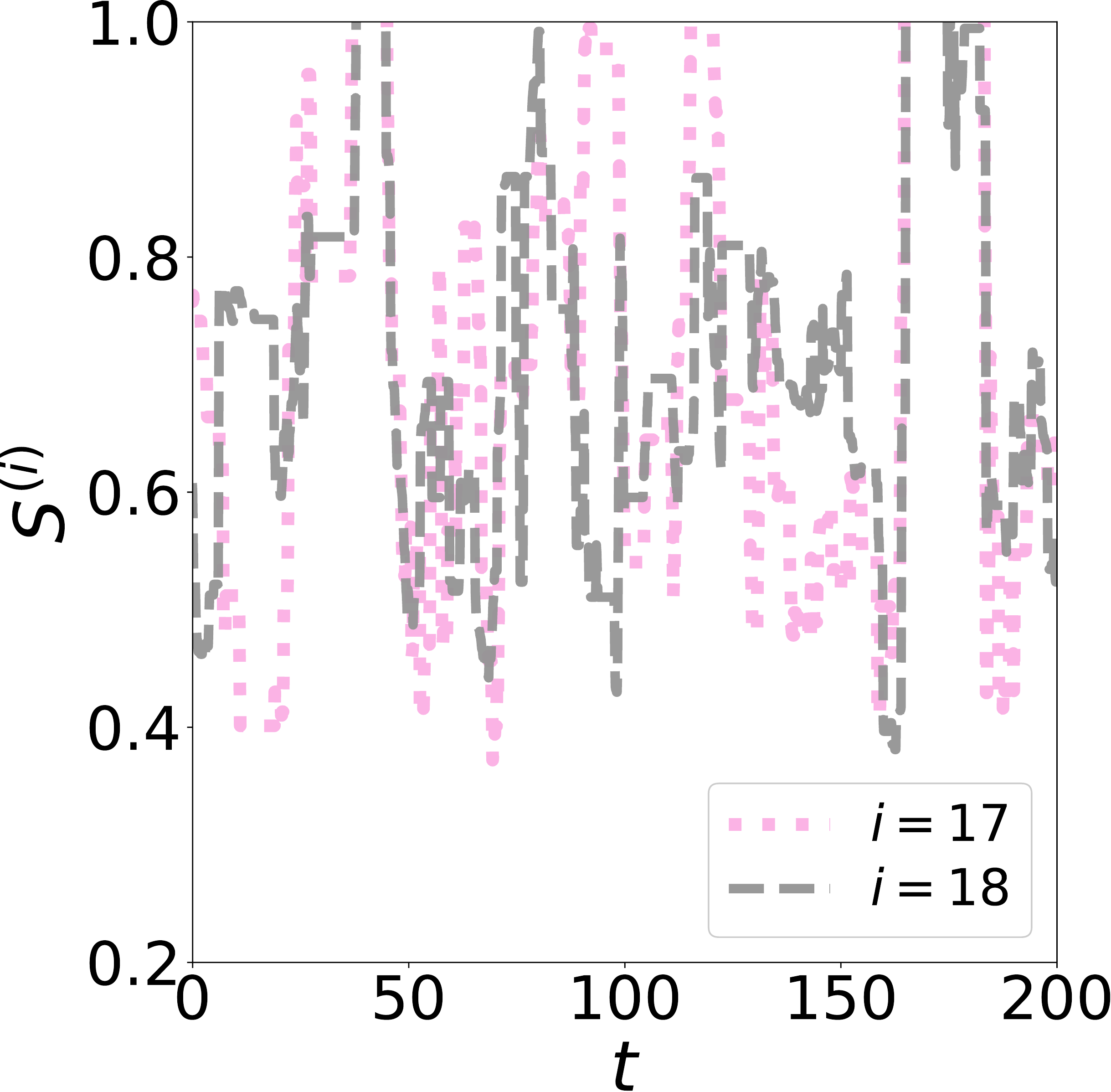}
        \put (0,0) {(c)}
    \end{overpic}
    \begin{overpic}[height=0.235\linewidth]{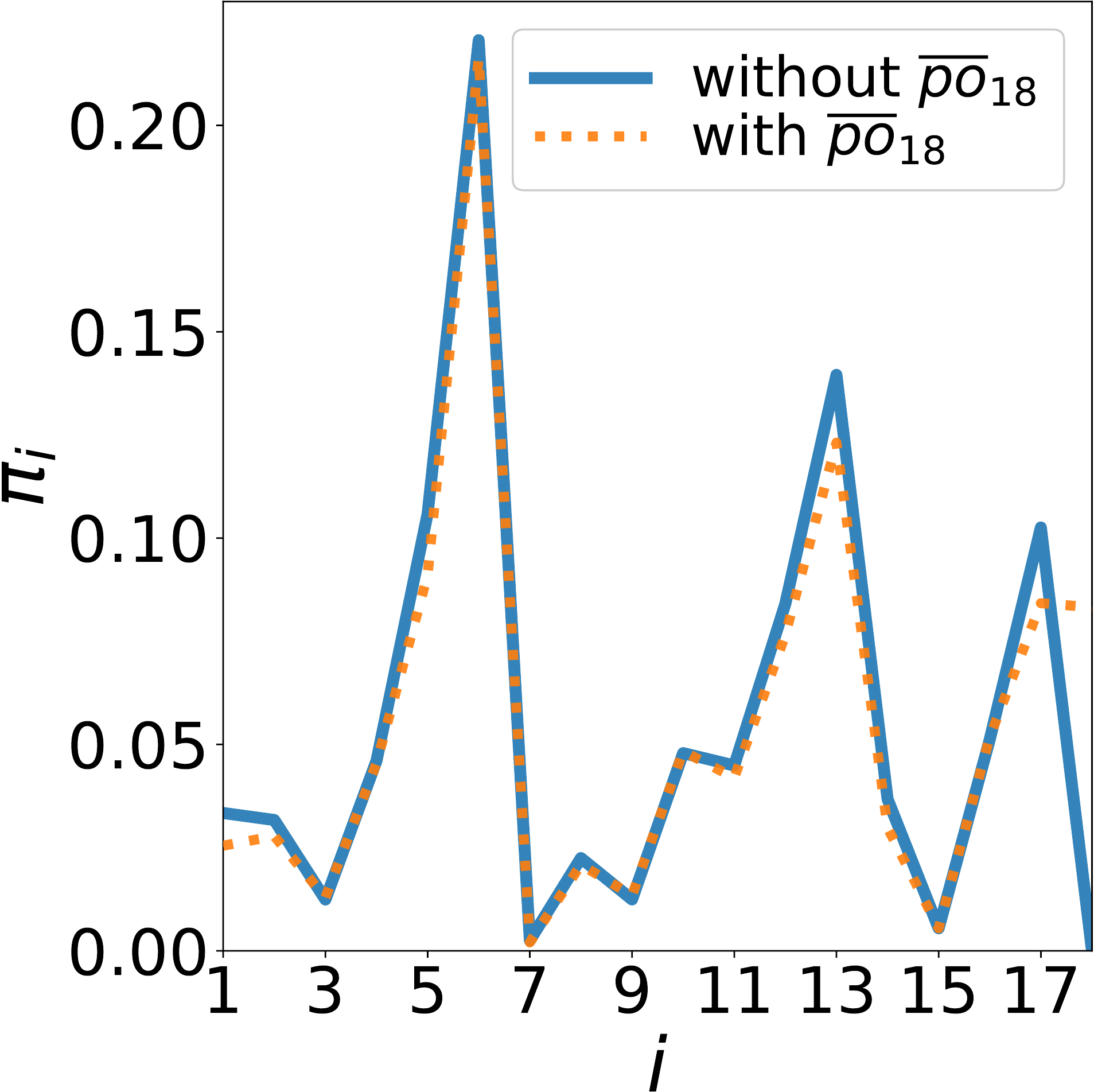}
    \put (0,0) {(d)}
    \end{overpic}
    \caption{
        (a) Heatmap showing the shadowing distances \refeq{e-distShadowPO} between 
        periodic orbits.
        (b) Projections of 
        \(\po_{17}\) (pink/thick) and 
        \(\po_{18}\) (gray/thin) 
        onto the first three
        principal components associated with \(\po_{17}\).
        (c) Shadowing distances $S^{(17,18)}(\zeit)$ of a 
        turbulent trajectory segment from $\po_{17}$ and $\po_{18}$.
        (d) Invariant distribution (\(\pi\)) computed with (blue/thick) 
        and without (yellow/dotted) \(\po_{18}\).
        The data points are connected with line segments for guiding 
        the eye.
        \label{f-shadowdistspairwise}}
\end{figure}

Since our recurrence-based periodic orbit search is an 
experimental process,
it is reasonable to expect some periodic orbits to 
be located nearby in the state space with similar physical 
properties. 
We search for such cases by defining a periodic orbit shadowing 
distance as follows. 
Let $\DataSet^{(i, \po_j)} =
\{\svecproj^{\po_j} (0),
  \svecproj^{\po_j} (t_s),
  \ldots,
  \svecproj^{\po_j} ((N_j - 1)t_s) \}$
be the projection of the states sampled on the $\po_j$ onto 
the bases of $\po_i$. We define the shadowing 
distance of $\po_j$ from $\po_i$ as 
\beq
	\distShadow^{(ij)} =
	w_0 W_{\infty} (\PerD^{(i, \po_j)}_0 (\zeit), \PerD^{(\po_i)}_0)
  + w_1 W_{\infty} (\PerD^{(i, \po_j)}_1 (\zeit), \PerD^{(\po_i)}_1) \,,
	\label{e-distShadowPO}
\eeq
where $\PerD^{(i, \po_j)}$ denotes the persistence diagrams 
associated with $\DataSet^{(i, \po_j)}$ and the weights 
$w_{0,1}$ are defined in the same way as in the shadowing 
distance of turbulence from $\po_i$. 
We visualized $\distShadow^{(ij)}$ as a heat map in 
\reffig{f-shadowdistspairwise}(a).
By definition \refeq{e-distShadowPO}, $\distShadow^{(ij)}$ is 
not symmetric under $i \rightleftarrows j$, which is also visible 
in \reffig{f-shadowdistspairwise}(a).
Nevertheless, if two periodic orbits $\po_i$ and $\po_j$ are located 
at nearby state space regions and possess similar shapes, we expect 
both $\distShadow^{(ij)}$ and $\distShadow^{(ji)}$ to be small. 
We found this to be the case for the pairs $(11,14)$ and 
$(17,18)$.
While $\po_{11}$ and $\po_{14}$ have considerably different 
$\langle E \rangle$ and $\langle D \rangle$, 
those of $\po_{17}$ and $\po_{18}$ agree in two digits. 
In order to illustrate their similarity, we visualized 
$\po_{17}$ and $\po_{18}$ as projections onto the principal components
associated with $\po_{17}$ in \reffig{f-shadowdistspairwise}(b). 
As a final test, we show the shadowing distances of a turbulent 
trajectory segment from $\po_{17}$ and $\po_{18}$ in
\reffig{f-shadowdistspairwise}(c),
where one can see 
that the minima of $S^{17}(\zeit)$ and $S^{18} (\zeit)$ always appear 
near one another. 
Based on these observations, we decided to retain only one of these 
periodic orbits in our model. 
We confirmed that this choice had no effect in the first two digits 
of the observable averages computed from the invariant distribution 
of our Markov chain.  
We show the invariant distributions of the Markov chains with and 
without $\po_{18}$ where one can see that the statistical weight of 
$\po_{18}$ is transferred predominantly to $\po_{17}$
when it is omitted.

\section{Spatial structures of the periodic orbits}
\label{s-spatial}
\begin{figure}[h]
        \begin{overpic}[height=0.25\linewidth]{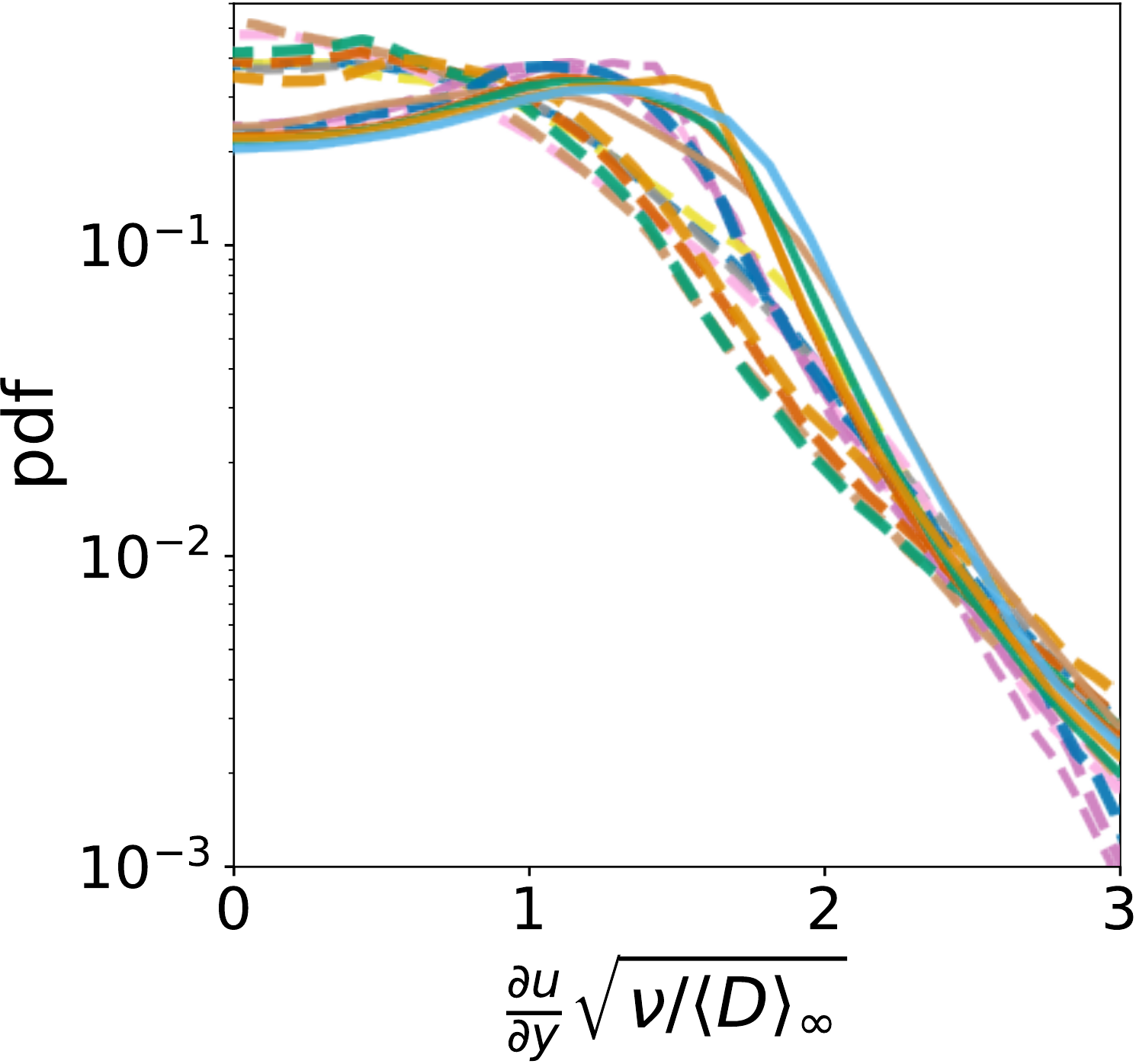}
        \put (0,0) {(a)}
        \end{overpic}
        \begin{overpic}[height=0.25\linewidth]{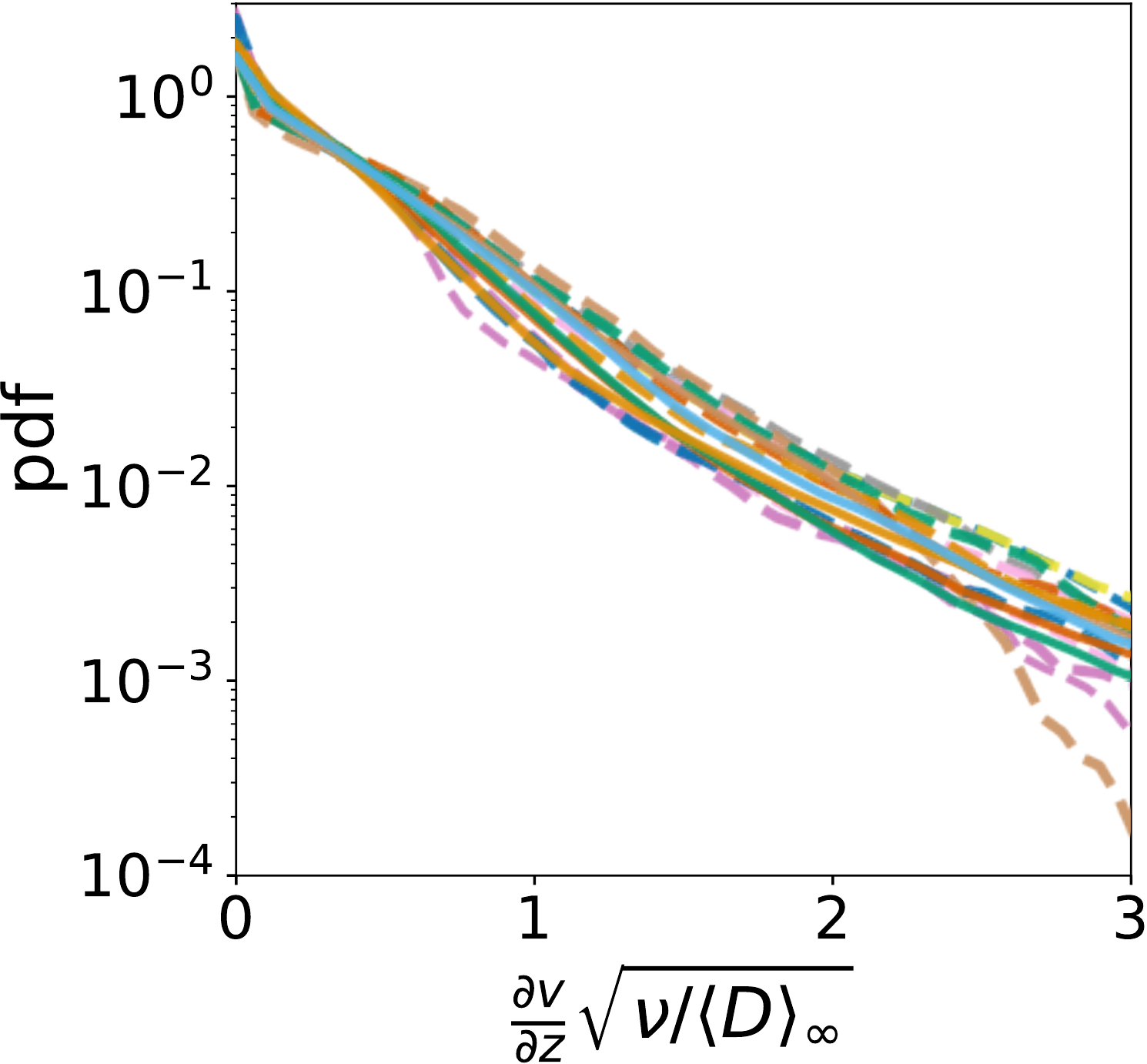}
            \put (0,0) {(b)}
        \end{overpic}
        \begin{overpic}[height=0.25\linewidth]{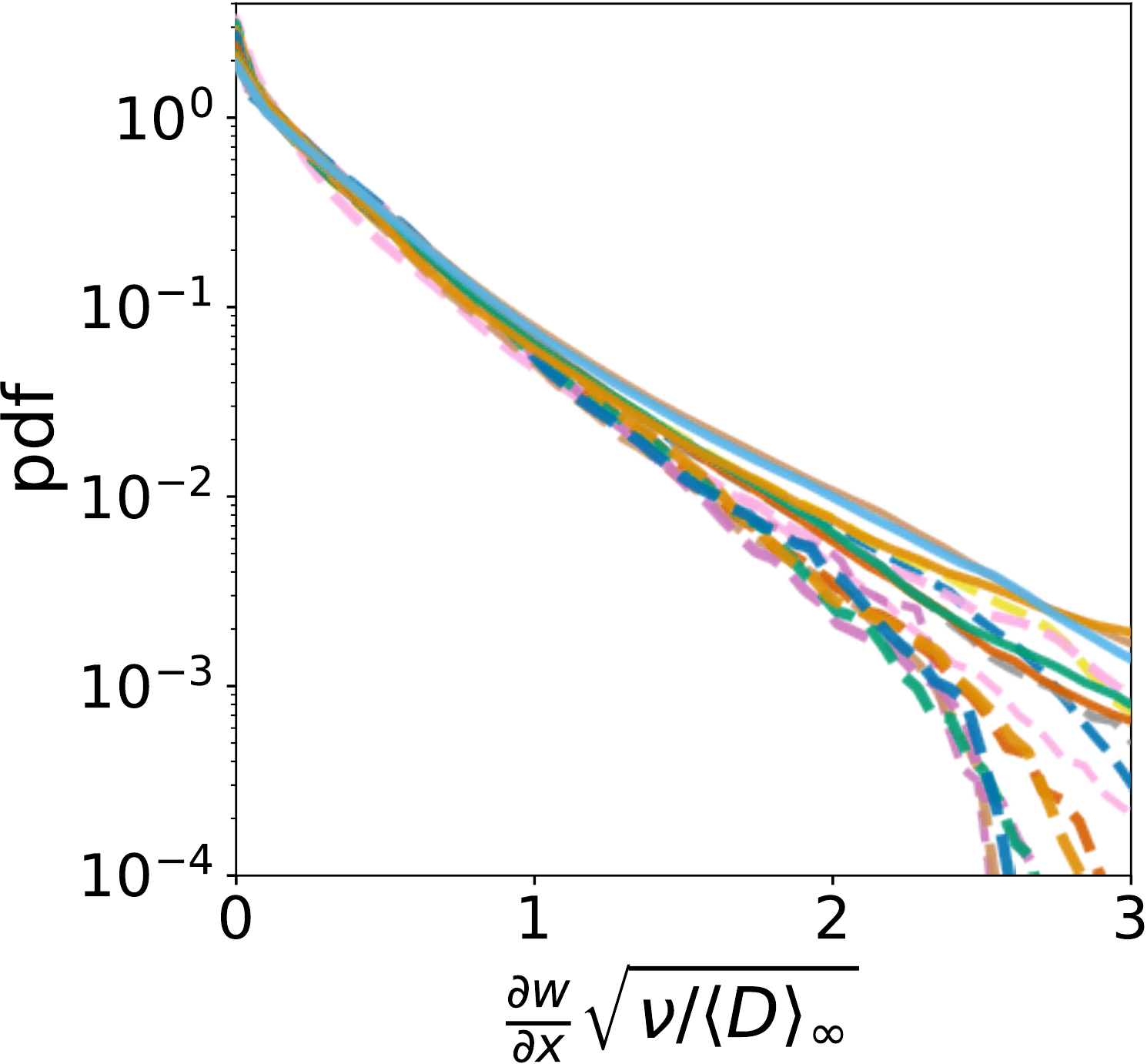}
            \put (0,0) {(c)}
        \end{overpic}\\
        \vspace{0.1cm}
        \begin{overpic}[height=0.25\linewidth]{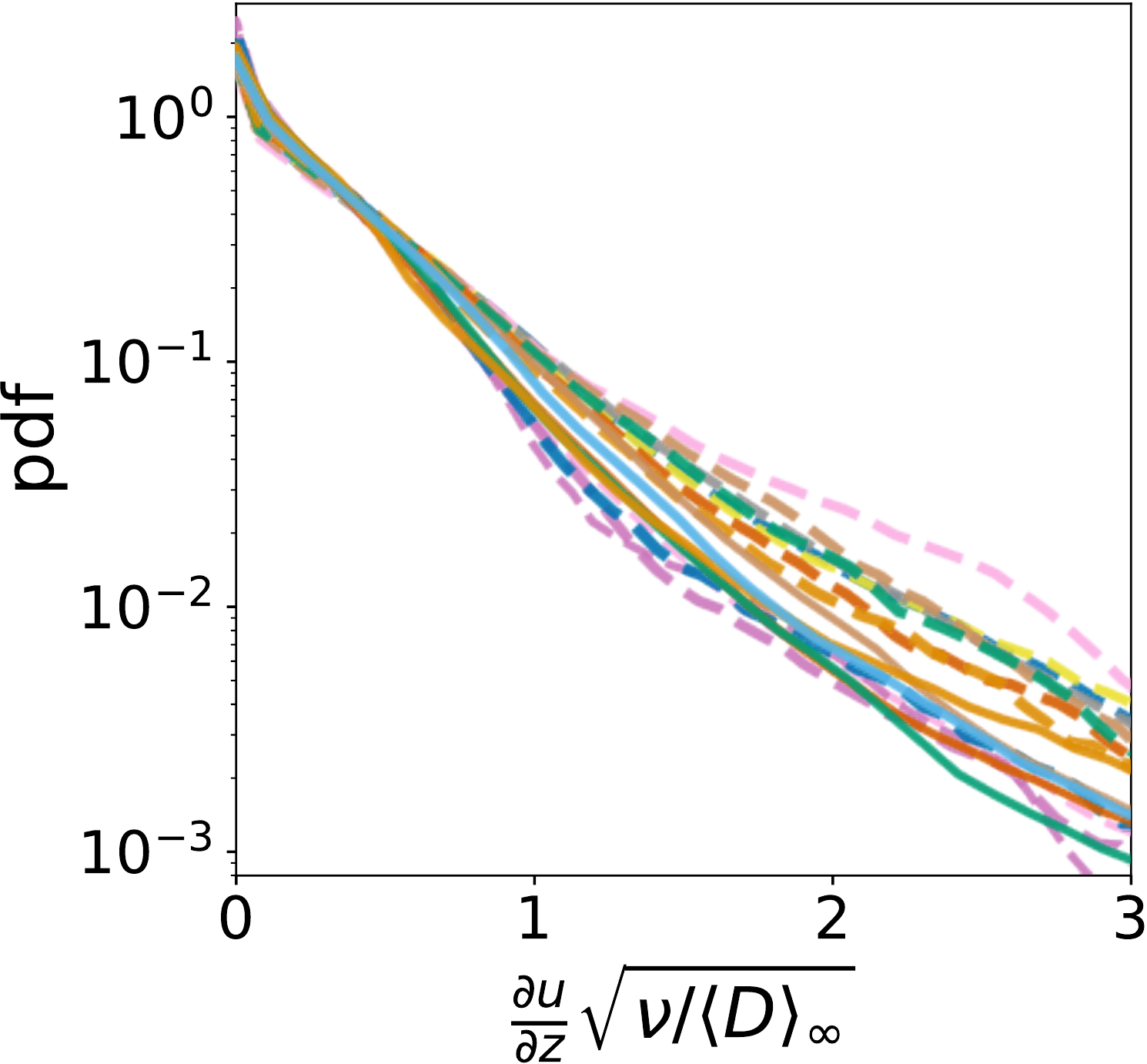}
            \put (0,0) {(d)}
        \end{overpic}
        \begin{overpic}[height=0.25\linewidth]{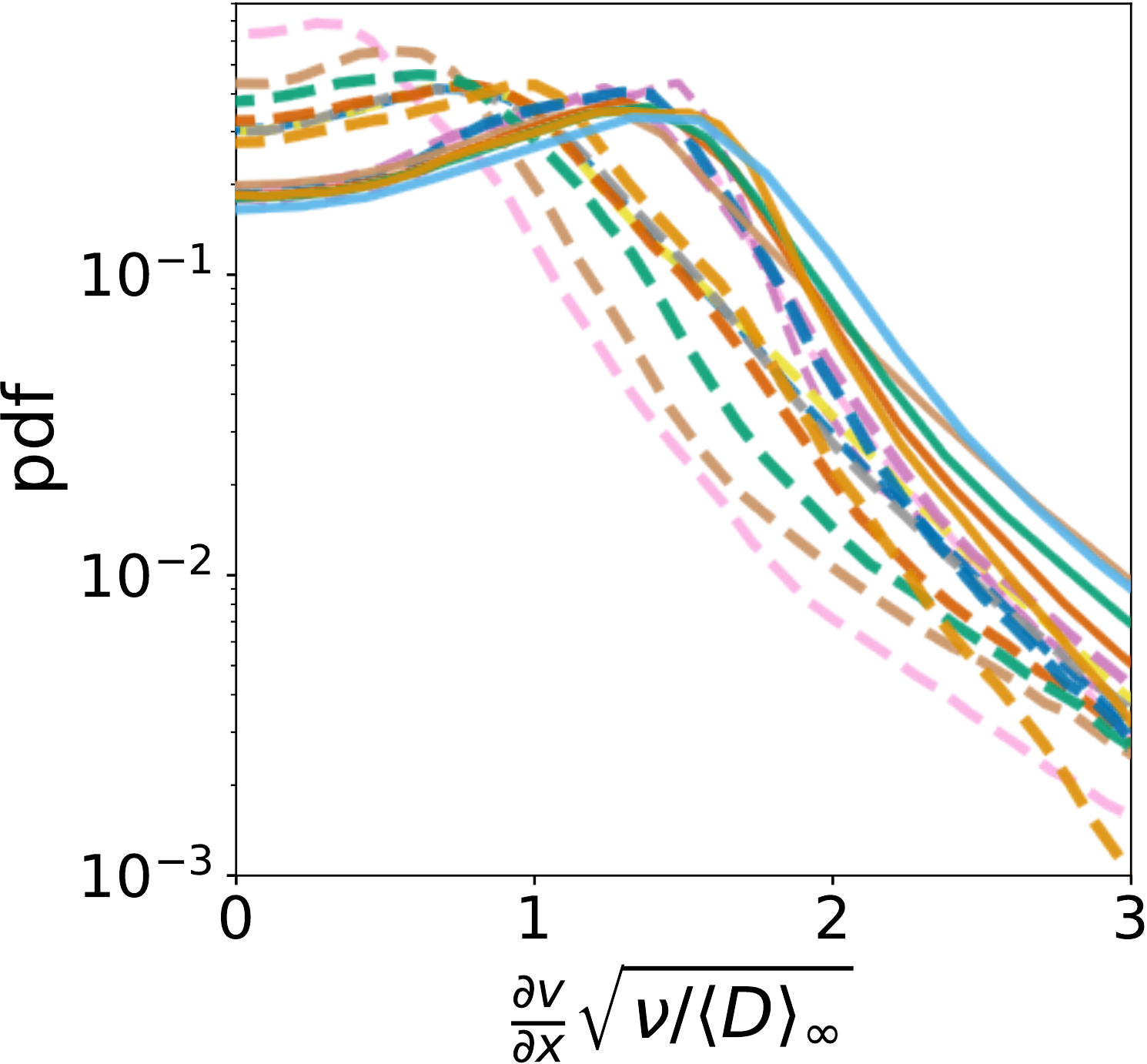}
            \put (0,0) {(e)}
        \end{overpic}
        \begin{overpic}[height=0.25\linewidth]{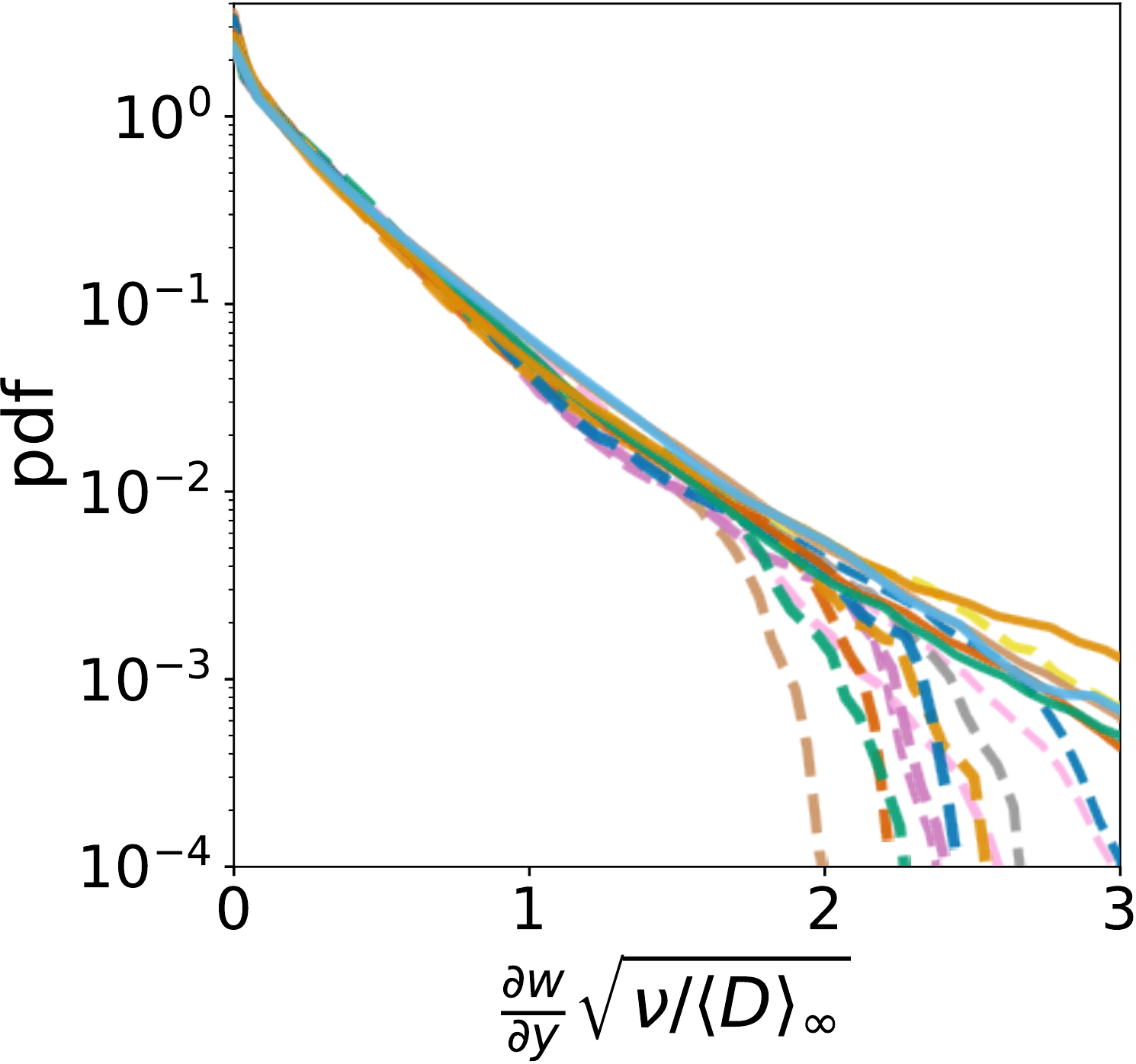}
            \put (0,0) {(f)}
        \end{overpic}
        \caption{
            Distributions of the gradients
            \(\partial u_i / \partial x_j,\, i \neq j\)
            on the periodic orbits.
            Only the positive half is shown since the distributions are symmetric 
            around \(0\).
            Solid (dashed) lines correspond to the periodic orbits with average dissipation
            greater (less) than \(\langle D \rangle_\infty\).
            \label{f-shadowgradients}}
\end{figure}

Distributions of velocity gradients are often of interest in turbulent 
flows where heavy tails imply increased energy dissipation. 
\reffig{f-shadowgradients} shows the distribution of gradients
\(\partial u_i / \partial x_j , i \neq j\) where the solid (dashed) lines correspond to 
\textpo{}s with average dissipation greater (less) than the long-time average of turbulence. 
As shown, the difference of distributions is most pronounced in 
$\partial u / \partial y$ (\reffig{f-shadowgradients}(a))
and $\partial v / \partial x$ (\reffig{f-shadowgradients}(e)) terms where
such solid 
curves appear to peak at higher values. 
    
\section{Partial shadowing of periodic orbits}

\begin{figure}[h]
    \begin{overpic}[height=0.21\linewidth]{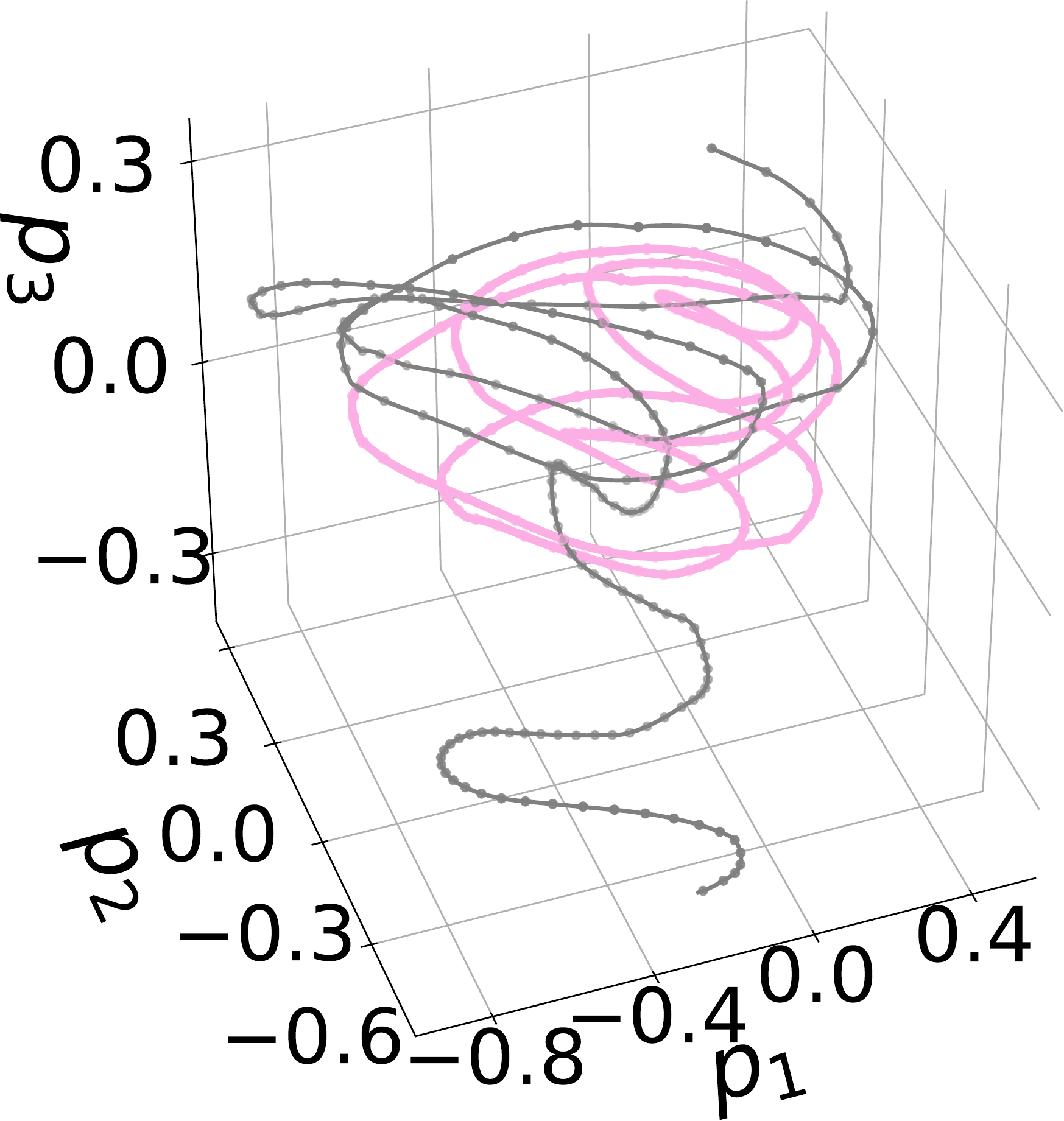}
        \put (0,0) {(a)}
    \end{overpic}
    \begin{overpic}[height=0.21\linewidth]{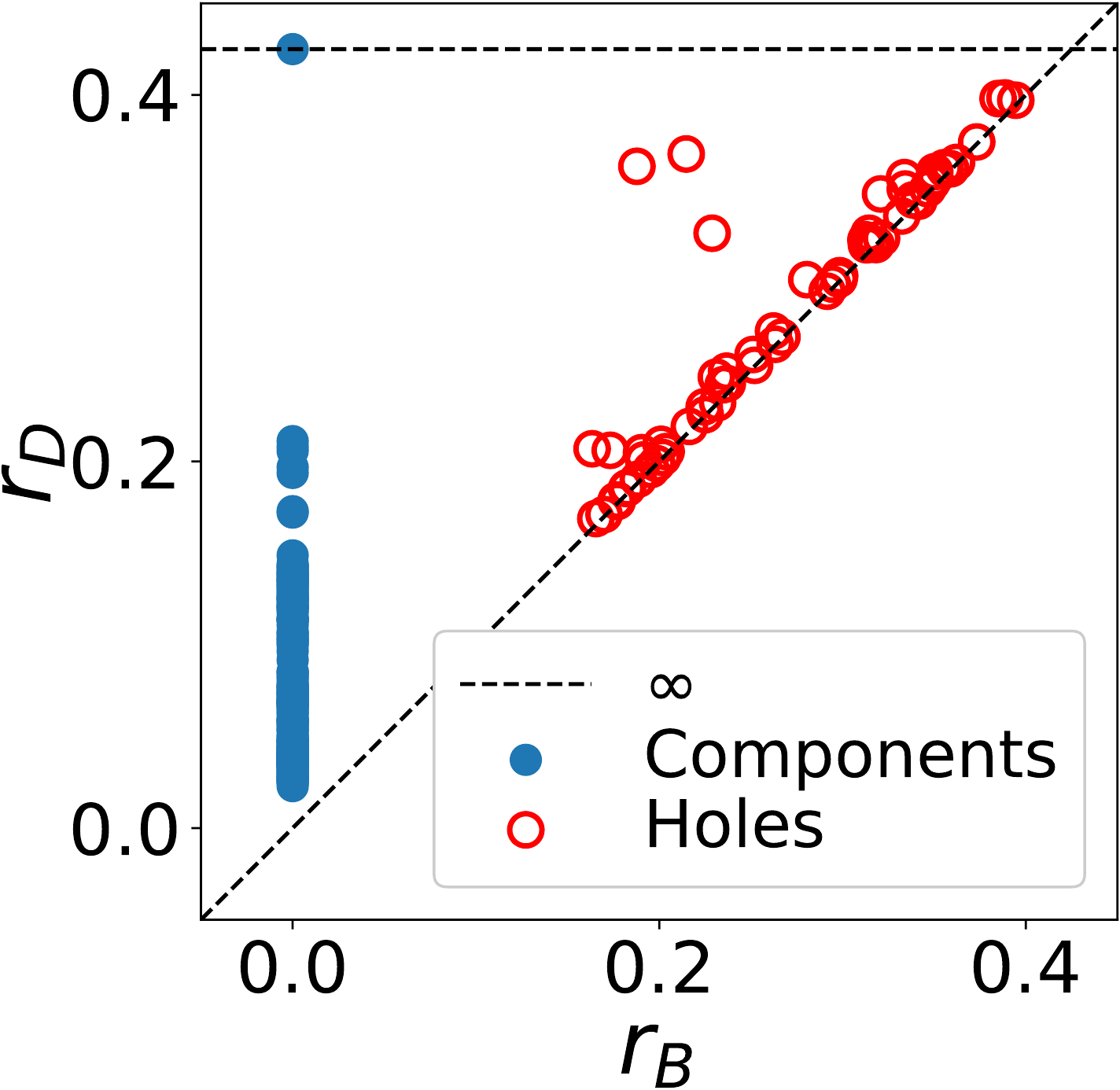}
        \put (0,0) {(b)}
    \end{overpic}
    \begin{overpic}[height=0.21\linewidth]{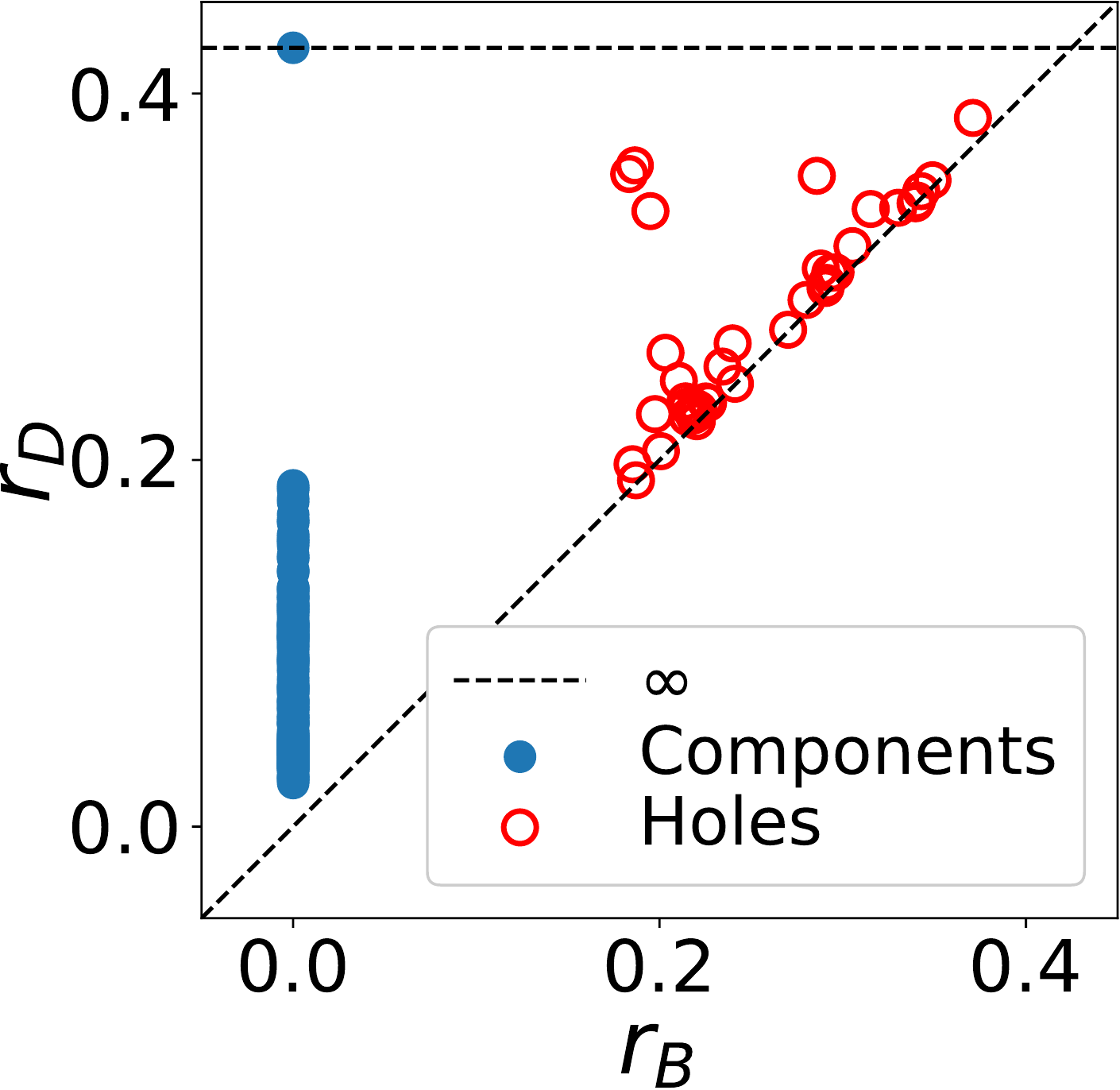}
        \put (0,0) {(c)}
    \end{overpic}    
    \begin{overpic}[height=0.21\linewidth]{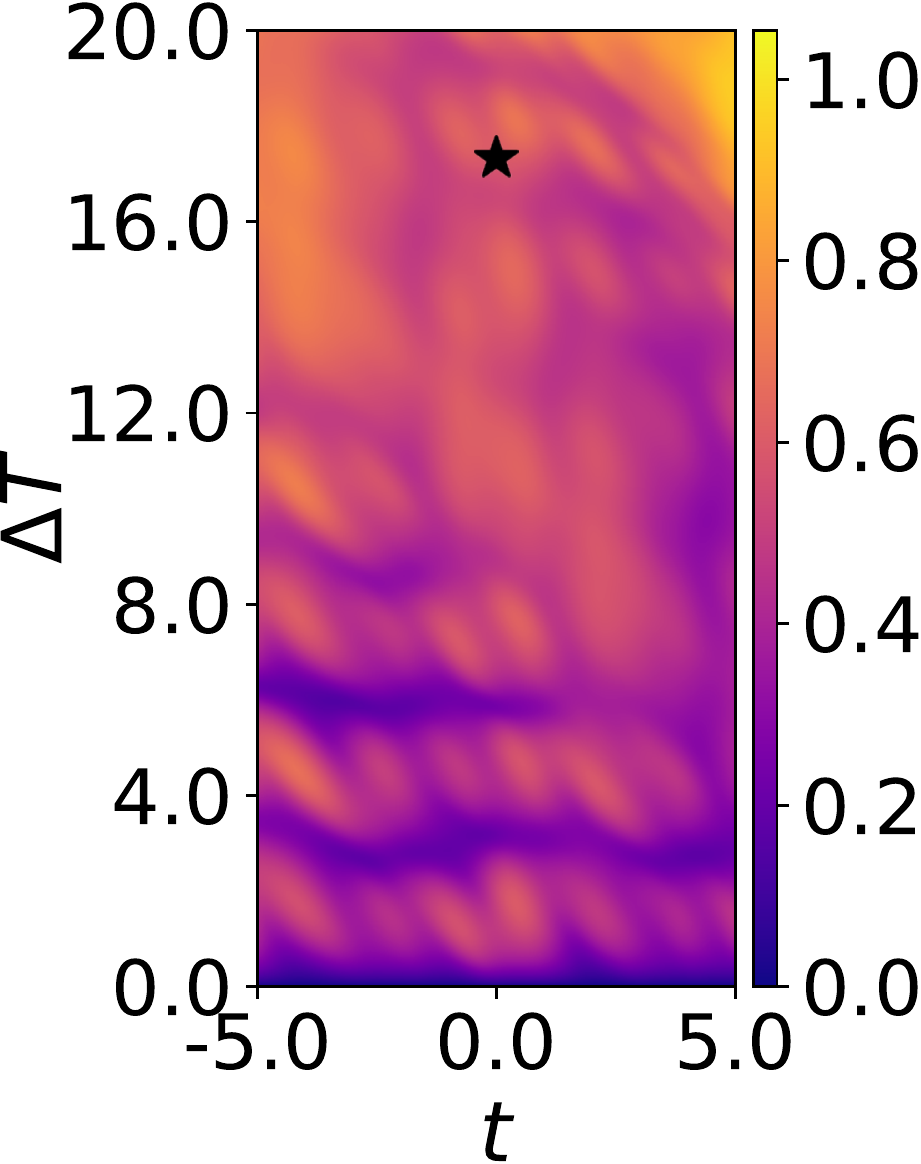}
        \put (0,0) {(d)}
    \end{overpic}    
    \begin{overpic}[height=0.21\linewidth]{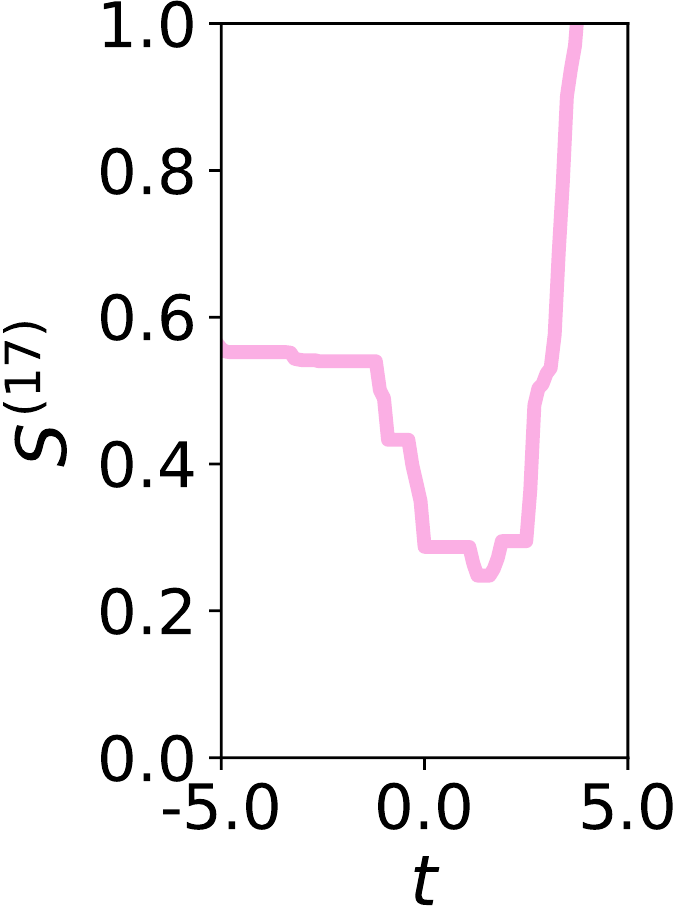}
        \put (0,0) {(e)}
    \end{overpic}    
    \caption{
        (a)
        Periodic orbit $\po_{17}$ (pink/thick)
        and a shadowing trajectory (gray/thin) visualized as
        projections onto the leading three principal components
        of $\po_{17}$.
        (b,c) The persistence diagrams associated with
        $\po_{17}$ (b) and shadowing trajectory segment (c)
        shown in (a).
        The data points used for generating the
        persistence diagrams (b,c) are marked with dots along
        the projection curves in (a).
        (d) Recurrence plot from a turbulent trajectory, where
        \(t=0\) corresponds to the state in the beginning of the trajectory
        shown in (a).
        The marked point (black star) would be the location of the respective
        recurrence signal, if this state closely recurred to itself after \(T_{17}\).
        (e) Shadowing distance time series corresponding to the same time interval
        in (a).        
        \label{f-proj-persist-episode1}
        }
\end{figure}

An important feature of our shadowing distance based on the shape 
similarity of turbulent trajectory segments and periodic orbits is 
its ability to detect shadowing events even when a turbulent trajectory 
follows only part of a periodic orbit. 
As an illustration, in \reffig{f-proj-persist-episode1} we show 
three-dimensional projections (\reffig{f-proj-persist-episode1}(a)) of
a periodic orbit and a turbulent trajectory segment along with the 
associated persistence diagrams (\reffig{f-proj-persist-episode1}(b,c)),
a recurrence plot (\reffig{f-proj-persist-episode1}(d)) and the corresponding
shadowing distance time series (\reffig{f-proj-persist-episode1}(e)).  
Clearly, the shadowing distance $\distShadow^{(17)} (\zeit)$ has a local 
minimum corresponding to this episode whereas the recurrence plot shows 
no signal at the period $T=17.3382$ of $\po_{17}$.
This episode is also visualized in the beginning of our supplementary 
video.

\section{Robustness against the choice of shadowing distance threshold}

\begin{figure}[h]
    \begin{overpic}[height=0.24\linewidth]{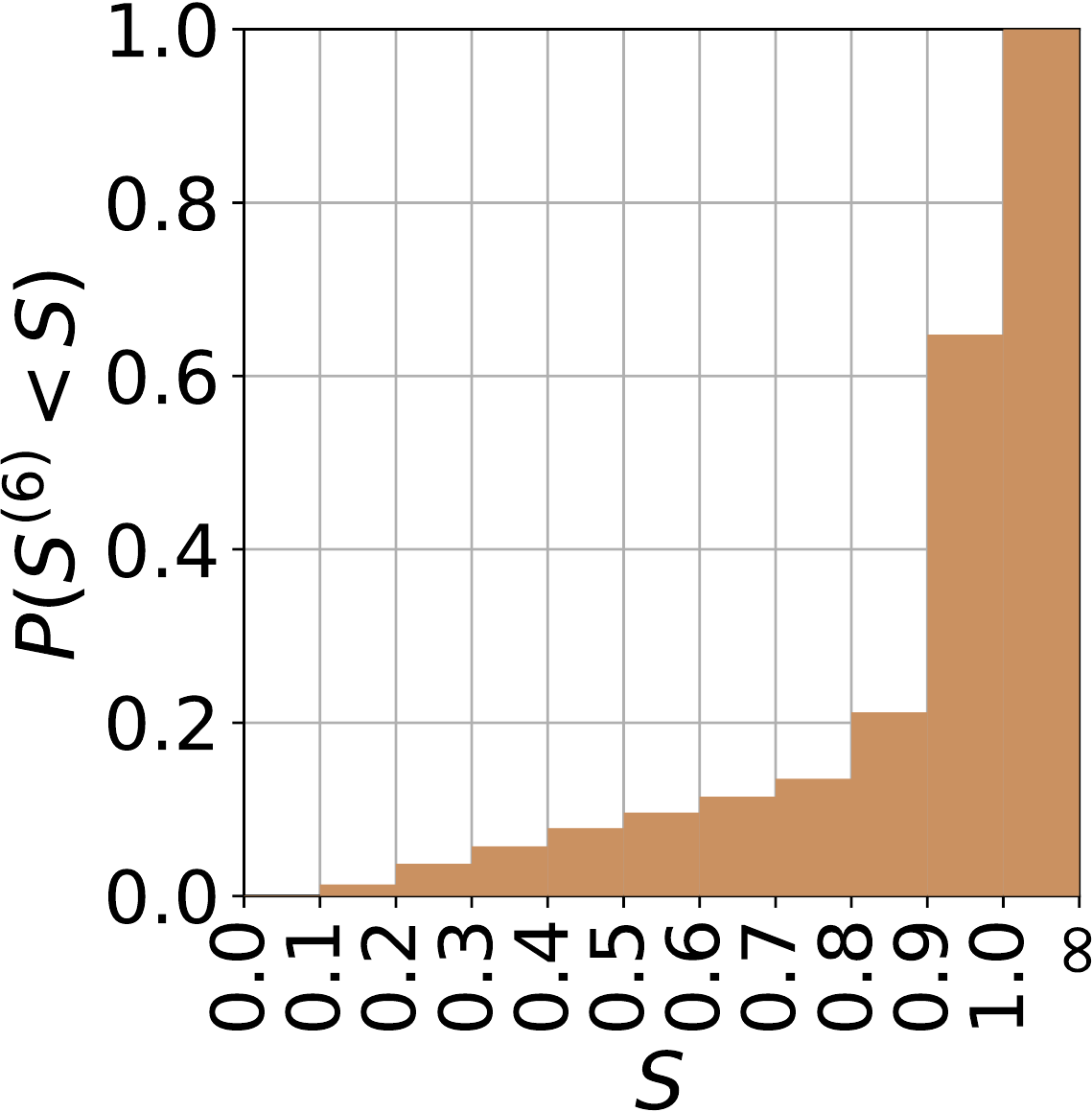}
        \put (0,0) {(a)}
    \end{overpic}    
    \begin{overpic}[height=0.24\linewidth]{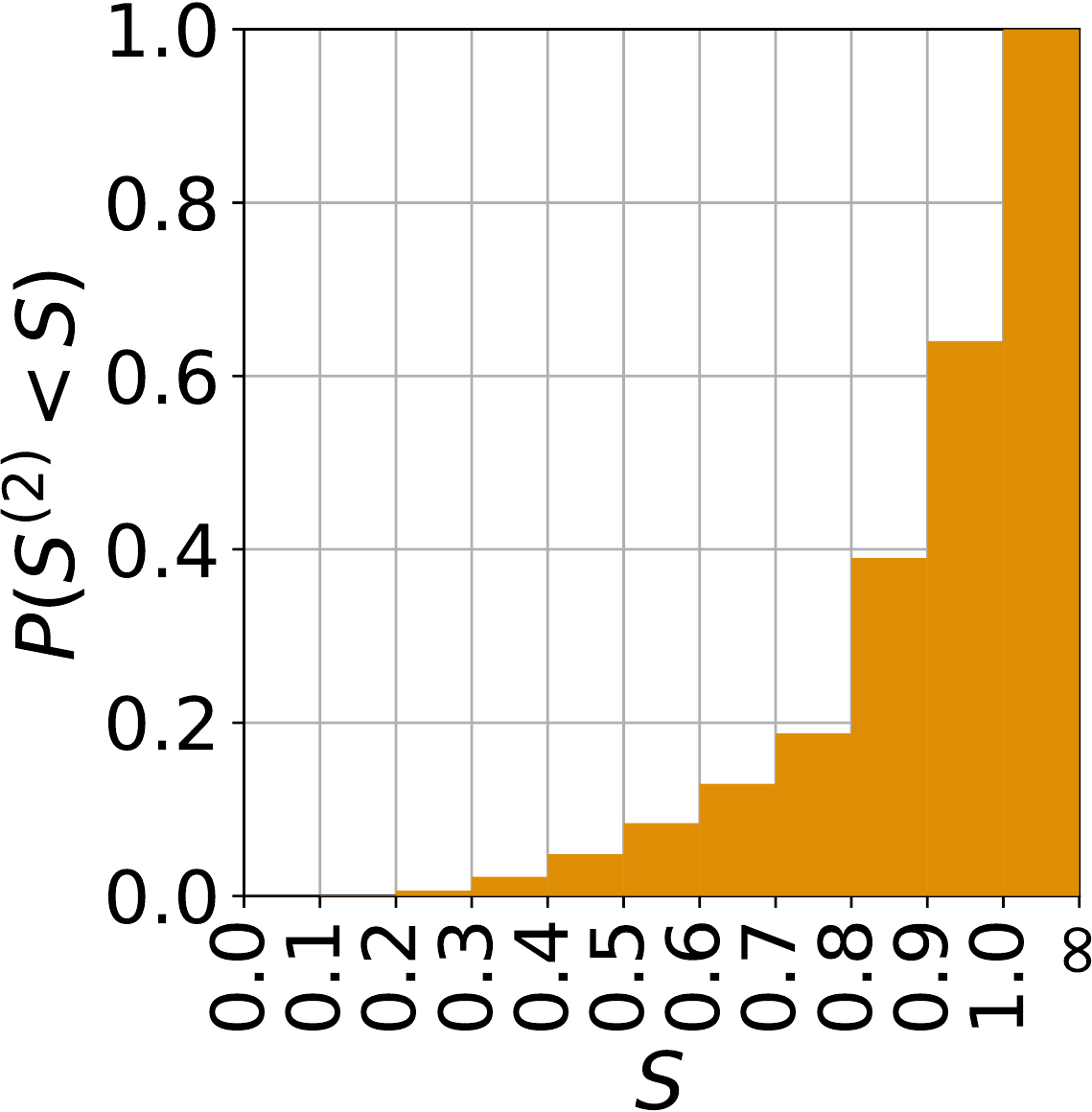}
        \put (0,0) {(b)}
    \end{overpic}    
    \begin{overpic}[height=0.24\linewidth]{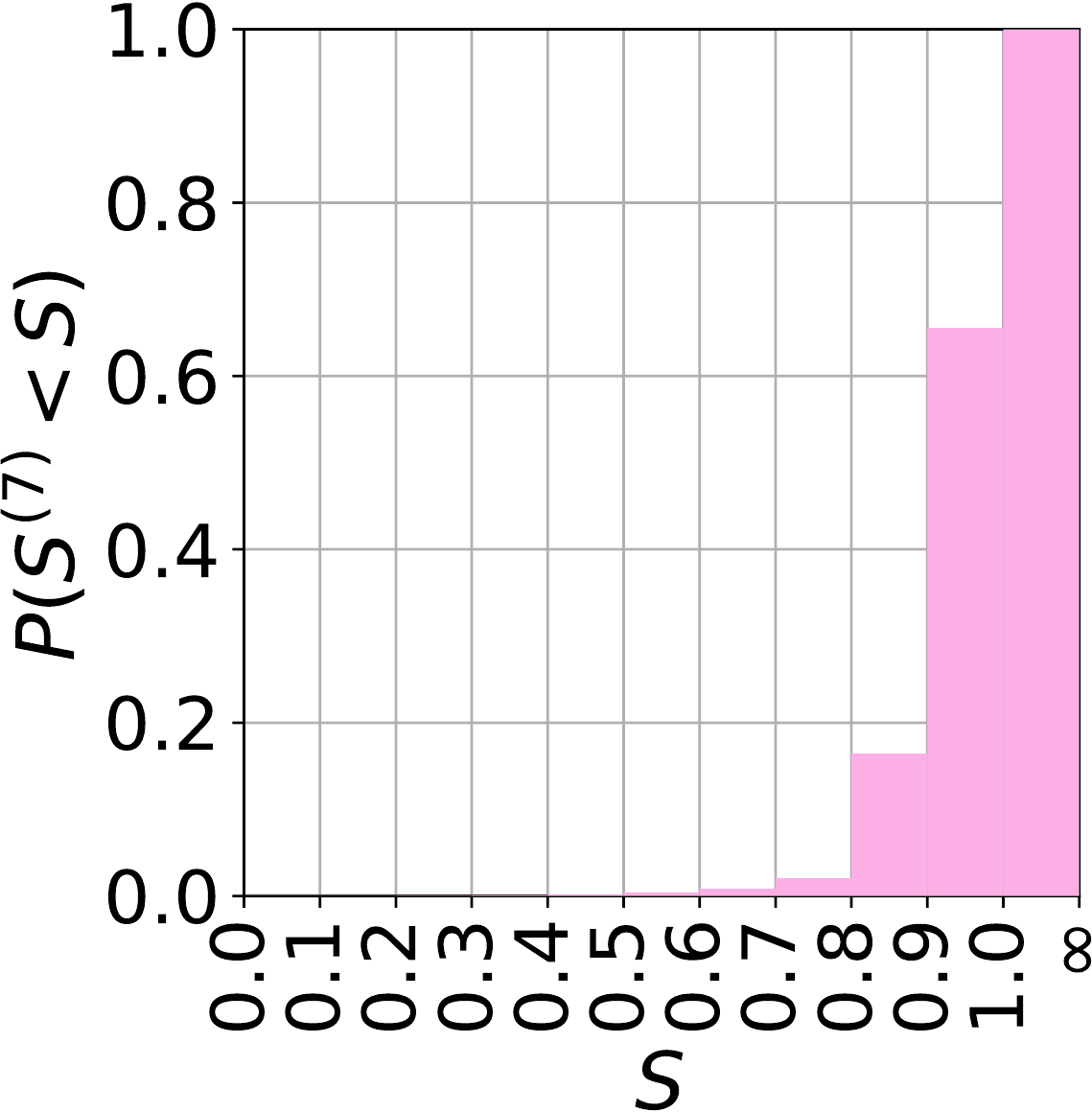}
        \put (0,0) {(c)}
    \end{overpic}    
    \begin{overpic}[height=0.24\linewidth]{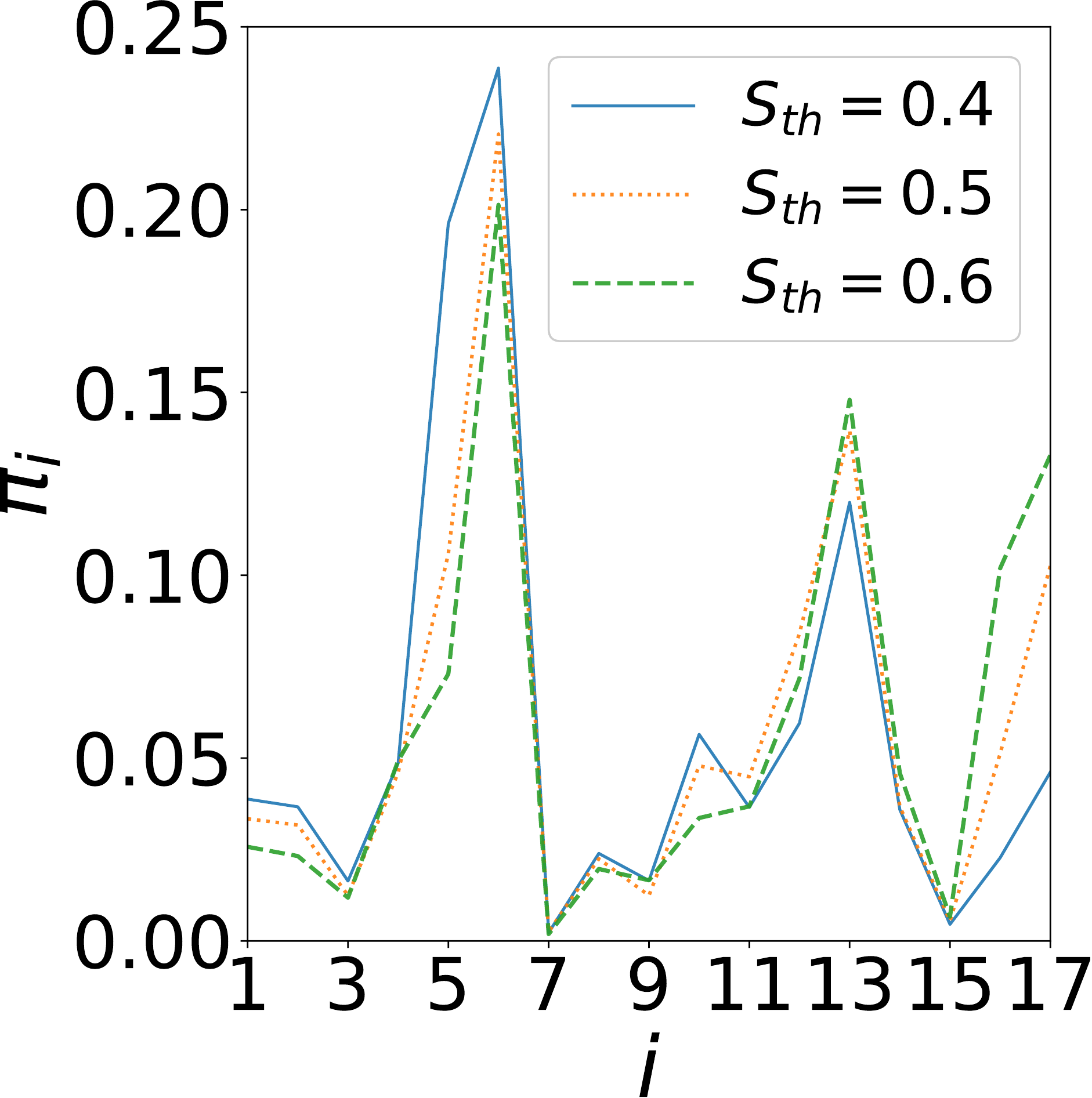}
        \put (0,0) {(d)}
    \end{overpic}    
    \caption{
        Cumulative distributions of the shadowing distance of 
        turbulence from 
        (a) $\po_6$,
        (b) $\po_2$, 
        and (c) $\po_7$.
        (d) Invariant distribution (\(\pi\)) computed using 
        different shadowing distance thresholds \(S_\text{th}\),
        with total run time \(t_\text{tot} = 25039\).
        The data points are connected with line segments for guiding 
        the eye.
        \label{f-SdistandThreshold}}
\end{figure}

As explained in the main text, a shadowing event is only registered when the 
shadowing distance of turbulence from a periodic orbit is less than the threshold 
$\distShadow_{th}$. 
Of course, our analysis can only be meaningful if our results do not strongly 
depend on the choice of this threshold. 
In \reffig{f-SdistandThreshold}(a--c) we show cumulative distributions 
of shadowing distances from $\po_6$, $\po_2$, and $\po_7$. 
We chose these as examples in order to illustrate cases corresponding 
to the highest ($\po_6$) and  lowest ($\po_7$)
statistical weights and an intermediate one ($\po_2$).
Even though the total amount of time that is decomposed into shadowing events 
varies as a function of our choice of $\distShadow_{th}$, 
the final invariant distribution changes only slightly 
for $S_{th} \in \{0.4, 0.5, 0.6\}$
as can be seen in \reffig{f-SdistandThreshold}(d). 

\FloatBarrier

\section{Convergence of statistics}

\begin{figure}[h]
    (a)\includegraphics[width=0.75\linewidth]{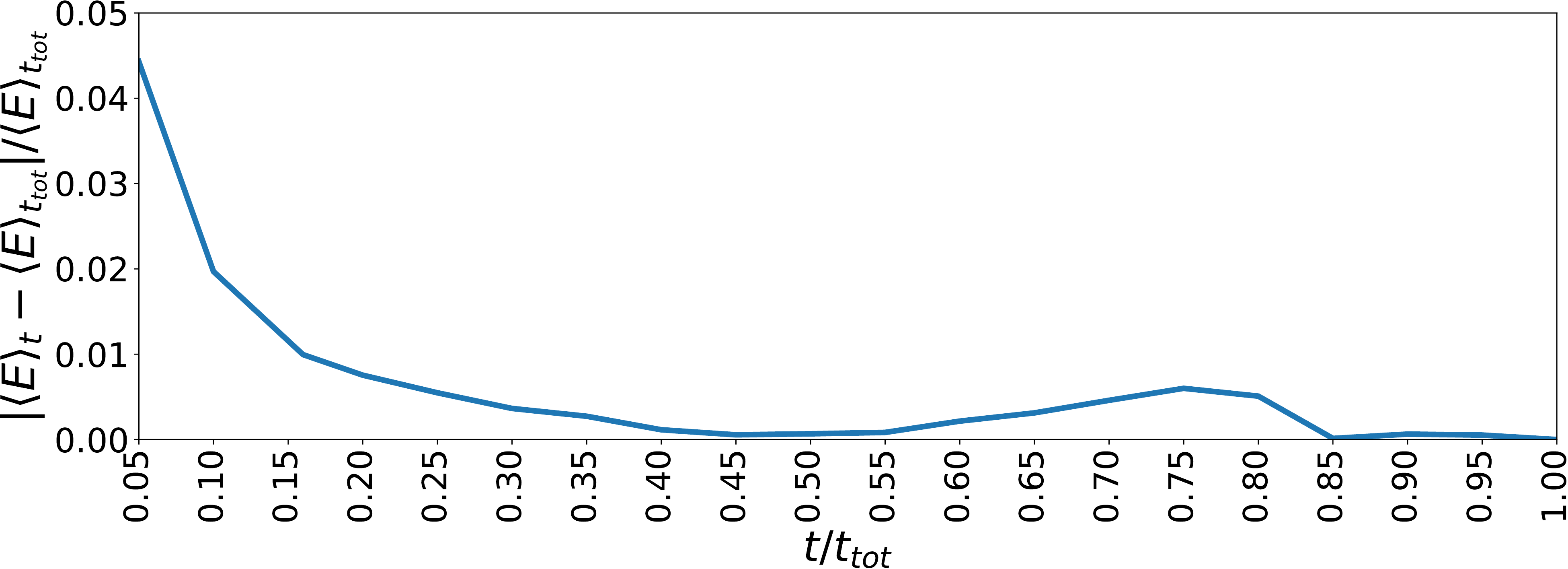}\\
    (b)\includegraphics[width=0.75\linewidth]{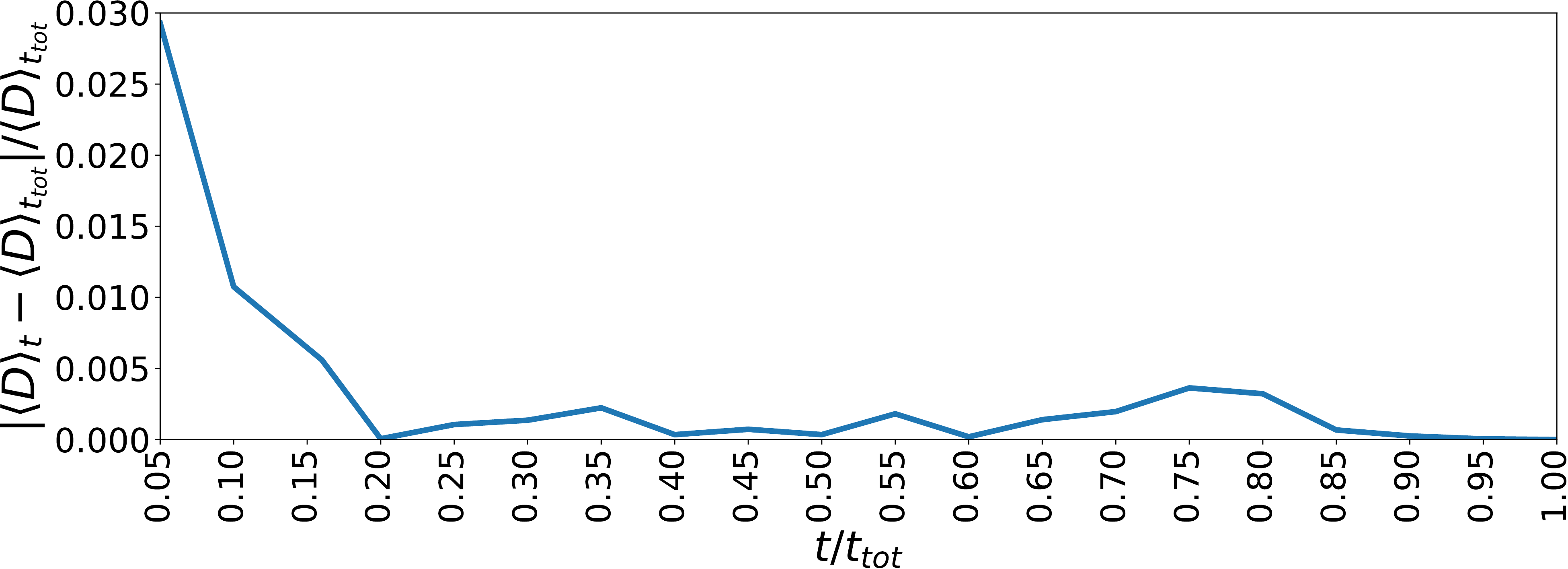}\\
    \caption{
        Relative errors of temporal averages of the 
        (a) kinetic energy and 
        (b) dissipation,
        computed over the 
        test dataset with a total lifetime $t_{tot} = 19214$.
        \label{f-turbEDconv}}
\end{figure}

In our analysis, we used three distinct data sets of turbulent dynamics:
(i) the recurrence set with the total runtime $t_\text{tot} = 5864$, 
(ii) the training set with the total runtime $t_\text{tot} = 25039$, 
and (iii) the test set with the total runtime $t_\text{tot} = 35492$.
The sole purpose of the test set was to compute the temporal averages of the 
kinetic energy and dissipation, convergence of which is shown in 
\reffig{f-turbEDconv}.
As shown, both averages remain within $1\%$ of their final value when 
more than $20\%$ of the data set is included. 

\begin{figure}[h]
    (a) \includegraphics[width=0.75\linewidth]{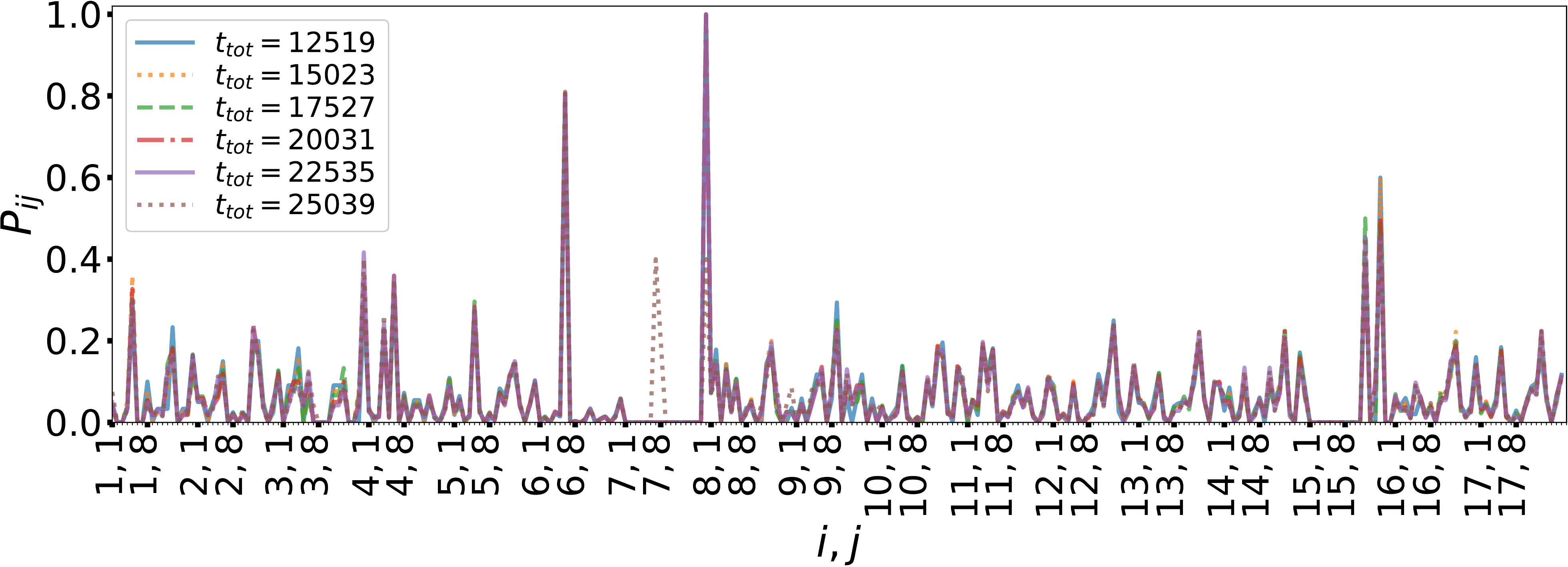}\\
    (b) \includegraphics[width=0.75\linewidth]{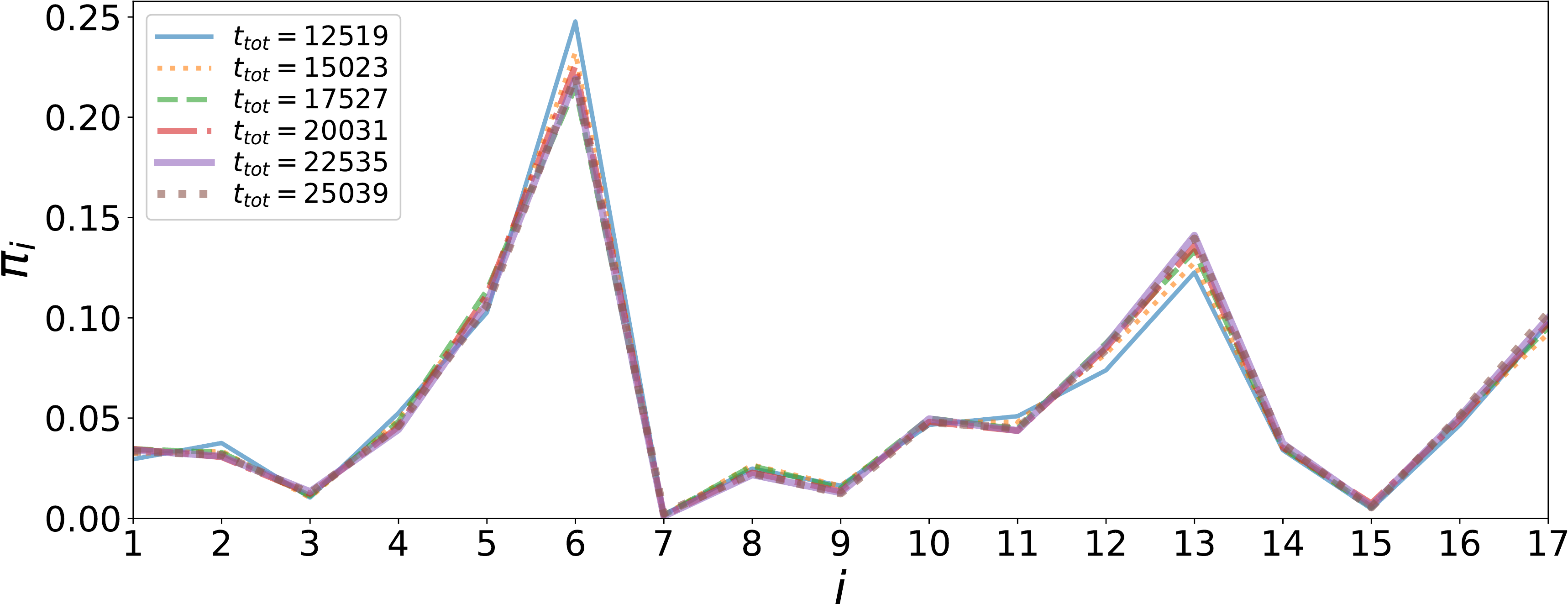}
    \caption{
        (a) Transition matrix (\(P\)) at different total
        run times (\(t_\text{tot}\)), with \(S_\text{th} = 0.5\).
        (b) Invariant distribution (\(\pi\)) at different total
        run times (\(t_\text{tot}\)), with \(S_\text{th} = 0.5\).
        The data points are connected with line segments for guiding 
        the eye.
        \label{f-transition-t}}
\end{figure}

We use the training data set for the shadowing decomposition of 
turbulence and, thus, inference of the transition matrix $\TransMat$. 
In \reffig{f-transition-t}(a), we visualized $\TransMat$ 
which we computed using data sets of varying total lengths.
In \reffig{f-transition-t}(a), the horizontal axis corresponds to the 
matrix entries ordered as pairs 
\[ [(1,1), (1,2), \ldots (1,17), (2,1), (2,2), \ldots (2,17), \ldots, 
  (17,1), (17, 2), \ldots (17, 17) ]. \]
In \reffig{f-transition-t}(b), we show the invariant distributions 
associated with each of these matrices.
While \reffig{f-transition-t}(a) appears to be mostly stable, a large 
peak at $i,j = 7, 8$ can be seen for $t_\text{tot} = 25039$ , which
is not present in the previous estimates. 
This large fluctuation is a consequence of the fact that 
$\po_7$ is not visited often by the turbulent flow, which can also 
be seen in the probability distribution of 
$S^{(7)}$ in \reffig{f-SdistandThreshold}(c). 
No such large fluctuation is visible in the invariant distributions 
shown in \reffig{f-transition-t}(b), suggesting that the long-time 
behavior inferred from the Markov chain is robust. 

\FloatBarrier

\begin{figure}[h]
    (a) \includegraphics[width=0.75\linewidth]{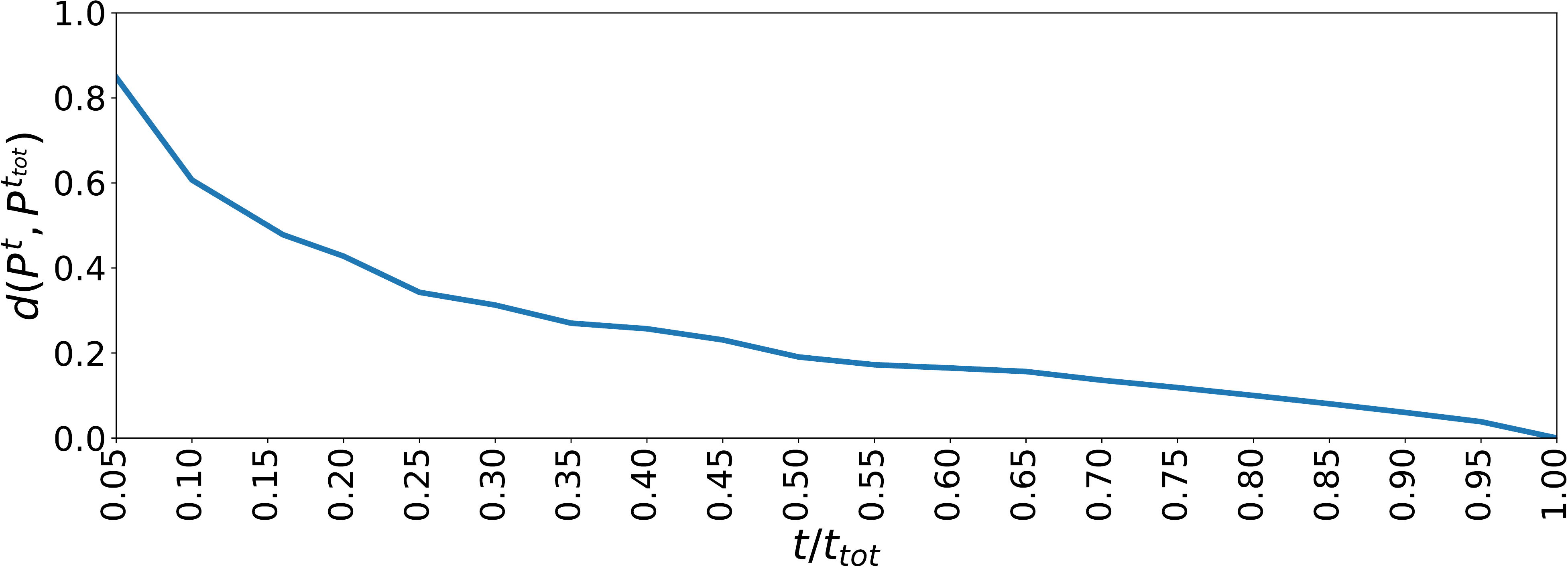}\\
    (b) \includegraphics[width=0.75\linewidth]{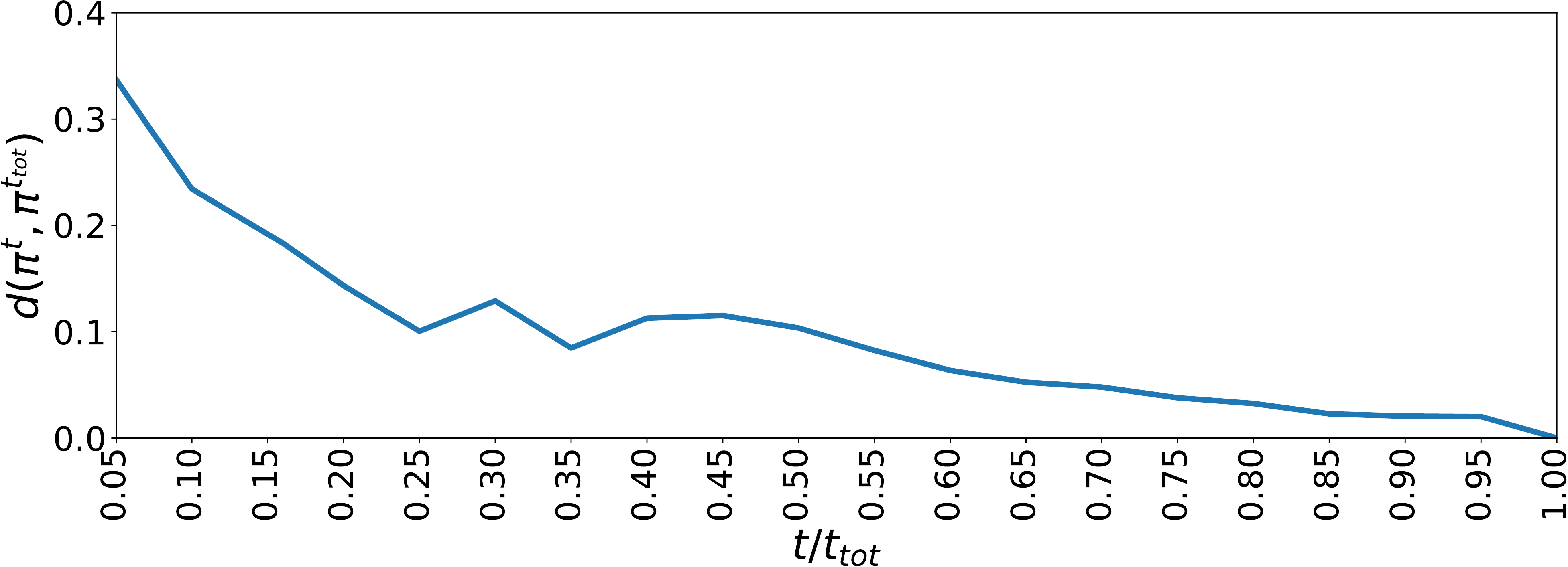}\\
    (c)\includegraphics[width=0.75\linewidth]{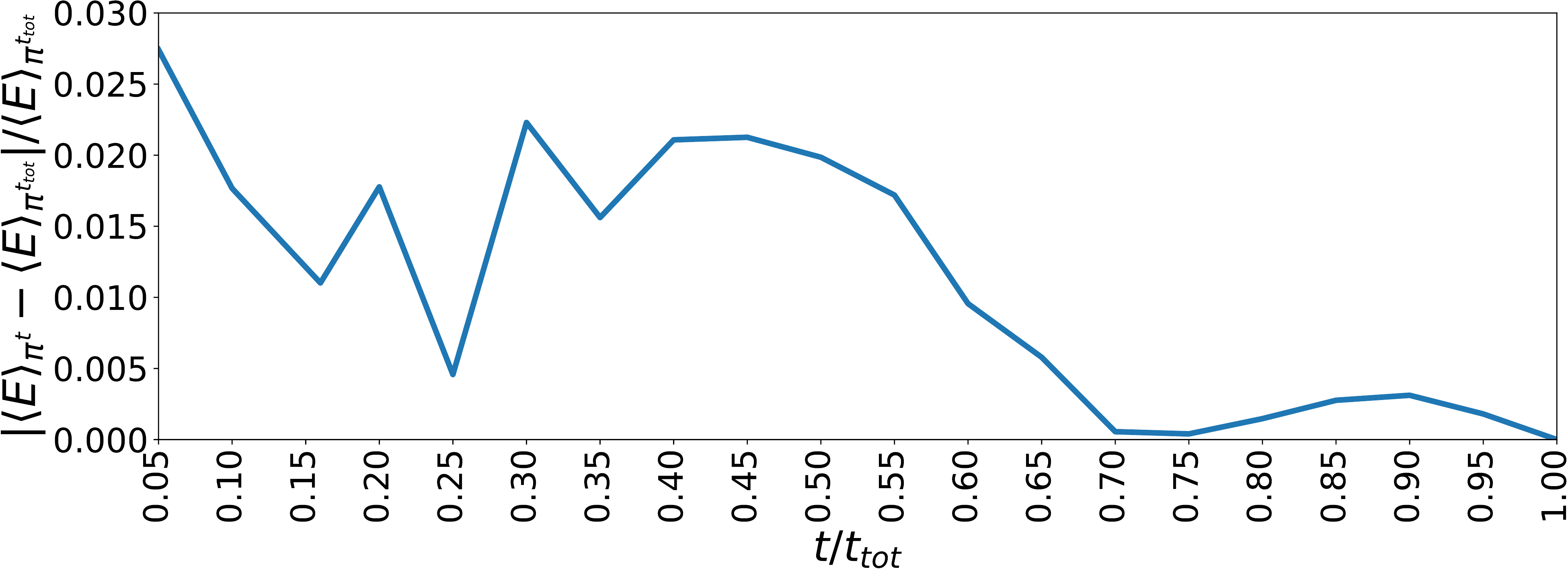}\\
    (d)\includegraphics[width=0.75\linewidth]{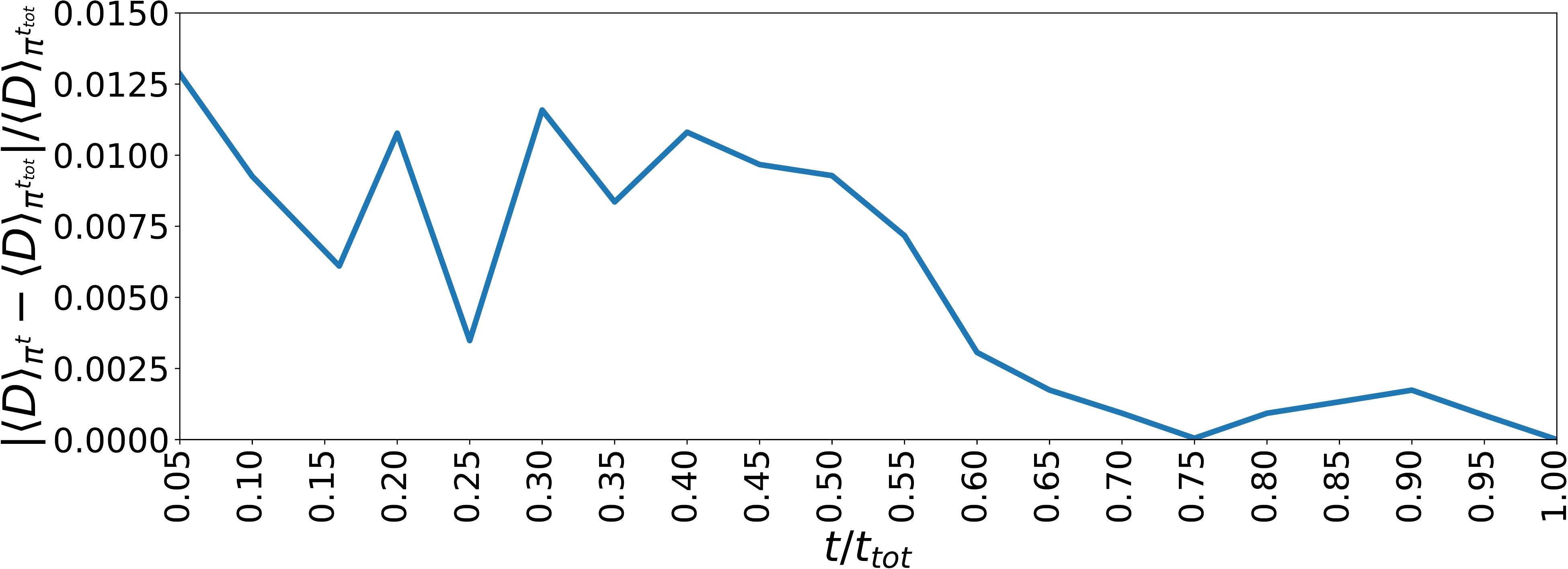}\\
    \caption{
        (a) Convergence of the transition matrix as measured by 
        \refeq{e-shadow-Pconv-weighted-frob}
        as the duration $\zeit$ of 
        training data set is increased.
        (b) Convergence of the invariant distribution as measured by 
        \refeq{e-shadow-dInvDist}
        as the duration $\zeit$ of 
        training data set is increased.
        (c,d) Convergence of the mean kinetic energy (c) and 
        dissipation (d)
        estimates over the invariant distributions $\InvDist^{t}$.
        \label{f-Pconv}}
\end{figure}

As another illustration of the convergence of our model, we show 
in \reffig{f-Pconv}(a) the deviation of $\TransMat^{\zeit}$ 
inferred from part of the training data set with duration $\zeit$
from its 
final estimate $\TransMat^{\zeit_\text{tot}}$  as measured by the metric
\begin{equation}
    d(\TransMat^{\zeit}, \TransMat^{\zeit_\text{tot}}) 
    =  \sum_{ij} \InvDist^{\zeit_\text{tot}}_i 
       \left|\TransMat^{\zeit}_{ij} - \TransMat^{\zeit_\text{tot}}_{ij}\right|\,,
    \label{e-shadow-Pconv-weighted-frob}
\end{equation}
where the sum is over all matrix entries. 
In \refeq{e-shadow-Pconv-weighted-frob}, weighing each row with its contribution 
to the final invariant measure emphasizes the node contributions accordingly, 
as well as sets $d(0, \TransMat^{\zeit_\text{tot}}) = 1$ where $0$ is the zero 
matrix.
Additionally, in \reffig{f-Pconv}(b), we show the convergence of the invariant 
distribution
$\InvDist^{\zeit}$ to its final estimate $\InvDist^{\zeit_\text{tot}}$
using the metric 
\begin{equation}
    d(\InvDist^{\zeit}, \InvDist^{\zeit_\text{tot}}) = 
    \sum_i\left|\InvDist^t_i - \InvDist^{t_\text{tot}}_i\right| \,.
    \label{e-shadow-dInvDist}
\end{equation}
Finally, in \reffig{f-Pconv}(c,d), we illustrate the convergence 
of the mean kinetic energy and dissipation, respectively, as measured by 
their relative error from the final estimates. 

\FloatBarrier

%% file: list-periodic-orbits.tex
\num{1} & \num{2.8076} & \num{10.6017} & \num{1.6967} & \num{3.3392e-02} \\
\num{2} & \num{2.9285} & \num{7.1407} & \num{1.5095} & \num{3.1696e-02} \\
\num{3} & \num{3.0481} & \num{5.2235} & \num{1.4172} & \num{1.2416e-02} \\
\num{4} & \num{3.2001} & \num{6.3834} & \num{1.4752} & \num{4.6056e-02} \\
\num{5} & \num{3.2027} & \num{10.4267} & \num{1.6435} & \num{1.0589e-01} \\
\num{6} & \num{3.3281} & \num{4.6274} & \num{1.3260} & \num{2.2066e-01} \\
\num{7} & \num{4.8178} & \num{4.2158} & \num{1.4045} & \num{2.5039e-03} \\
\num{8} & \num{5.7962} & \num{7.1860} & \num{1.6826} & \num{2.2472e-02} \\
\num{9} & \num{5.9469} & \num{7.5110} & \num{1.7262} & \num{1.2433e-02} \\
\num{10} & \num{11.1259} & \num{14.2693} & \num{2.3542} & \num{4.7897e-02} \\
\num{11} & \num{11.9670} & \num{7.1459} & \num{1.6708} & \num{4.4907e-02} \\
\num{12} & \num{14.0560} & \num{12.2741} & \num{2.0303} & \num{8.4089e-02} \\
\num{13} & \num{14.8255} & \num{12.7118} & \num{2.0253} & \num{1.3957e-01} \\
\num{14} & \num{15.0668} & \num{12.0885} & \num{1.9465} & \num{3.6768e-02} \\
\num{15} & \num{15.2772} & \num{11.1163} & \num{1.6996} & \num{5.4522e-03} \\
\num{16} & \num{16.5225} & \num{11.9875} & \num{2.1675} & \num{5.1218e-02} \\
\num{17} & \num{17.3382} & \num{10.6745} & \num{1.7239} & \num{1.0258e-01} \\
\num{18} & \num{17.0106} & \num{10.7257} & \num{1.7930} & \num{0} \\